\documentclass[twocolumn]{raa}            

\usepackage{graphicx,times}             
\usepackage{natbib}
\usepackage{multirow}
\usepackage{amssymb,amsmath}
\usepackage{textcomp}
\bibpunct{(}{)}{;}{a}{}{,}

\usepackage[a4paper=true,driverfallback=dvipdfm,pagebackref=true]{hyperref}

\hypersetup{colorlinks = true, linkcolor = green, anchorcolor = red, citecolor = blue, filecolor = red, pagecolor = red, urlcolor = red}

\begin{document}

   \title{The Quasar Candidates Catalogs of DESI Legacy Imaging Survey Data Release 9
}

   \volnopage{Vol.0 (20xx) No.0, 000--000}      
   \setcounter{page}{1}          

   \author{Zizhao He
        \inst{1,2}
        \and Nan Li
        \inst{1}
    }

	\institute{Key lab of Space Astronomy and Technology, National Astronomical Observatories,            Beijing, 100101, China;
		{\it nan.li@nao.cas.cn}
		\and
		School of Astronomy and Space Science, University of Chinese Academy of Sciences, Beijing 100049, China;\\
		\vs\no
		{\small Received 20xx month day; accepted 20xx month day}
	}
\abstract{Quasars can be used to measure baryon acoustic oscillations at high redshift, which are considered as direct tracers of the most distant large-scale structures in the Universe. It is fundamental to select quasars from observations before implementing the above research. This work focuses on creating a catalog of quasar candidates based on photometric data to provide primary priors for further object classification with spectroscopic data in the future, such as The Dark Energy Spectroscopic Instrument (DESI) Survey.  We adopt a machine learning algorithm (Random Forest, RF) for quasar identification. The training set includes $651,073$ positives and $1,227,172$ negatives, in which the photometric information are from DESI Legacy Imaging Surveys (DESI-LIS) \& Wide-field Infrared Survey Explore (WISE), and the labels are from a database of spectroscopically confirmed quasars based on Sloan Digital Sky Survey (SDSS) and the Set of Identifications \& Measurements and Bibliography for Astronomical Data (SIMBAD). The trained RF model is applied to point-like sources in DESI-LIS Data Release 9. To quantify the classifier's performance, we also inject a testing set into the to-be-applied data. Eventually, we obtained $1,953,932$ Grade-A quasar candidates and $22,486, 884$ Grade-B quasar candidates out of $425,540,269$ sources ($\sim 5.7\%$). The catalog covers $\sim 99\%$ of quasars in the to-be-applied data by evaluating the completeness of the classification on the testing set. The statistical properties of the candidates agree with that given by the method of color-cut selection. Our catalog can intensely decrease the workload for confirming quasars with the upcoming DESI data by eliminating enormous non-quasars but remaining high completeness.  All data in this paper is publicly available online.
\keywords{quasars: general --- catalogs --- methods: statistical}
 }

   \authorrunning{Z.He \& N.Li}            
   \titlerunning{The Quasar Candidates Catalogs of DESI-LIS DR9}  

   \maketitle

\section{Introduction}
\label{sect:intro}

    The discovery of quasars, also known as quasi-stellar objects (QSOs), is one of the four significant findings that have been made in astronomy in the 60s of last century \citep{Schmidt1963, Kellermann2014}. QSOs are extremely luminous active galactic nuclei  \citep[][AGN]{Osterbrock1989, Urry1995, Dunlop2003, Croton2006} powered by accretion onto supermassive black holes at the centers of galaxies, and their typical luminosity is ${10}^{42}$ to ${10}^{48}$ erg/s \citep{Shen2020} at the redshift from 0.1 $\sim 7$ \citep{Antonucci1993}. The emission of QSOs can significantly outshine their host galaxies, and their emitting regions are too small to resolve even for the nearest ones. Hence, QSOs are always considered point-like sources, which mimic faint blue stars in optical bands. However, they are $\sim 2$ magnitudes brighter in the near-infrared at all redshifts than stars of similar optical magnitudes and colors, leading to a neat way to discriminate QSOs from stars \citep{Ross2012,Myers2015,Yeche2020}.

    The nature of QSOs has been investigated widely and thoroughly in the past decades by using tons of corresponding observations. Consequently, QSOs are used to study astrophysical problems in various fields. For instance, the spectrum of QSOs is a powerful tracer of the formation and evolution of black holes, the spins of black holes, and the co-evolution of black holes and host galaxies \citep{Kormendy1995, Silk1998, Kaspi2000, Matteo2005, Springel2005, Kormendy2013, Chen2021, Valentini2021}; taking advantage of microlensing, astronomers study the feature of accretion disks with the light curves of QSOs \citep{Agol1999, Morgan2010, Dexter2011, Blackburne2011}; the absorption lines of quasars are unique tracers of the interstellar media along the line of sight \citep{Scaringi2009, Hall2013, Chen2020, Mishra2021, Zabl2021}. Besides, high-redshift quasars are valuable for understanding the reionization of the Universe and the formation of galaxies \citep{McLure2002, Wang2019, Lupi2021}. Statistically, the spatial distribution quasars reflects the baryon acoustic oscillations \citep[BAOs, e.g. ][for an introduction]{Zhao2019} and in turn the large-scale structure of the Universe \citep{Dawson2013,Ribera2014, Delubac2015,Alam2021, Merz2021}. Expectedly, with the next-generation large-scale surveys coming, an unprecedented dataset of QSOs brings an unparalleled opportunity to trigger a revolution in these fields. 

    Mining QSOs from enormous datasets is crucial for carrying out the studies mentioned above, and plenty of progress has been made. The Palomar-Green Bright Quasar Survey \citep[BQS,][]{Schmidt1983} discovered more than $100$ quasars. The Large Bright Quasar Survey \citep[LBQS,][]{Hewett1995} discovered more than $1,000$s. The 2-degree Field Quasar Redshift Survey \citep[2QZ,][]{Croom2004} discovered about $23,000$s. The Large Sky Area Multi-object Fiber Spectroscopic Telescope \citep[LAMOST,][]{Dong2018} discovered more than $20,000$s. At the moment, the largest confirmed quasar catalog is from Sloan Digital Sky Survey Data Release $16$ \citep[SDSS DR16,][]{Blanton2017,Lyke2020}, which contains $750,414$ spectrally confirmed quasars. Nevertheless, to implement confirming QSOs with spectrums, one must create a sample of QSO candidates using or even combining various types of data other than spectroscopy. For instance, photometry data can be used to select QSO candidates according to the features of QSOs, such as the ultraviolet excess, infrared excess, and the light variation \citep{Shen2011}; astrometry data kicks out the objects with high proper motion in the Milky Way \citep{Fu2021}; and, of course, Radio and X-ray data are valuable complements \citep{Bisogni2021}. 

    The Dark Energy Spectroscopic Instrument \citep[DESI,\footnote{\url{https://www.desi.lbl.gov/}}][]{Levi2013,DESICollaboration2016} is a spectral telescope that located at Kitt Peak National Observatory (KNPO). It is a Mayall telescope with a four-meter-aperture primary mirror. It will target about $30$ million pre-selected galaxies across $\sim 14,000$ square degree sky. It is important for the discovery of more quasar because of its large sky coverage, good image quality and depth \citep[compared to SDSS,][]{York2000}, and because it can provide spectrum. However, the QSO candidates is needed for further conforming the QSOs. Hence, we acquire the QSO candidates from the photometry catalog of DESI Legacy Imaging Survey (DESI-LIS).

    In this work, to create a catalog of QSO candidates for DESI, we adopt a machine learning (ML) technology named Random Forest (RF) and apply it to photometry data from DESI-LIS because its efficiency, flexibility, and accuracy have been intensively proved previously \citep[e.g. ][]{Viquar2018, Bai2019, Clarke2020, Guarneri2021}, in particular, \citet{Bai2019} demonstrates that RF is the most efficient and reliable one among several methods in dealing with quasar-star-galaxy classification. The training and validation sets are built upon the spectra data from SDSS eBOSS \citep[extended Baryon Oscillation Spectroscopic Survey, ][]{Dawson2016} DR16 and  the photometry data from WISE\footnote{\url{https://irsa.ipac.caltech.edu/Missions/wise.html}} \citep[Wide-field Infrared Survey Explorer,][]{Wright2010} and DESI-LIS, labels are generated based on the database of SIMBAD\footnote{\url{http://simbad.u-strasbg.fr/}} \citep[the Set of Identifications, Measurements and Bibliography for Astronomical Data,][]{Wenger2000}. To evaluate the completeness, accuracy, and purity of identifying QSOs candidates from the photometry data of point-like sources in DESI-LIS and WISE, we inject a testing set that mimics to-be-applied data in magnitude and color space. Later, the trained model is applied to point-like sources in DESI-LIS, and the quasar candidate catalogs are acquired. At last, we compare our results to those of the color-cut selection approach for cross-validation, and they match well. For the convenience of other researchers, we make the all the data in this paper publicly available online\footnote{\url{https://github.com/EigenHermit/he-li2021}}.

    The paper is organized as follows. The construction of the datasets used in this paper is presented in Sec. \ref{sec:data}. Sec. \ref{sec:mthd} introduces the details of the methods for detecting QSOs adopted in this study. We then show the results in Sec. \ref{sec:cnclsn}. At last, Sec. \ref{sec:dscssn} delivers the discussion and conclusions. In this paper, a fiducial cosmological model $\Omega_m=0.26$, $\Omega_{DE}=0.74$, $h=0.72$, $w_0=-1$ and $w_a=0$ is adopted. The cosmology is the same as the one adopted in \citet[][OM10 hereafter]{OM10}.

\section{Data-sets} 
\label{sec:data}

The datasets adopted in this work include DESI-LIS, WISE, SDSS eBOSS DR16, and SIMBAD. The training set combines photometry data from DESI-LIS and WISE, while the labels are from the confirmed QSO catalog from SDSS eBOSS DR16 and SIMBAD. The validation set is extracted from training set. To evaluate the performance of the classification of QSO candidates, we also build a testing set with the SIMBAD database and eBOSS dataset. Introduction to the above datasets and details of the construction of training and testing sets are described below.

\subsection{DESI-LIS,WISE,eBOSS,SIMBAD} \label{subsec:AvlblObsrvtns}
    
    \quad \ \ \ \textbf{DESI Legacy Imaging Surveys and WISE}

    DESI Legacy Imaging Surveys\footnote{\url{https://www.legacysurvey.org/}} \citep[][DESI-LIS]{Dey2019} contains Dark Energy Camera Legacy Surveys (DECaLS\footnote{\url{https://www.legacysurvey.org/decamls/}}), Beijing-Arizona Sky Survey (BASS\footnote{\url{https://www.legacysurvey.org/bass/}}) and Mayall $z$-band Legacy Survey (MzLS\footnote{\url{https://www.legacysurvey.org/mzls/}}), covering $\sim$ $14,000$ deg$^2$ of the extra-galactic sky in three optical bands ($g$, $r$, and $z$). Notebly, DESI-LIS DR9 also includes four mid-infrared bands (at $3.4$, $4.6$, $12$, and $22$ $\mu m$, corresponding to $W1$, $W2$, $W3$ and $W4$ repectively) observed by WISE\footnote{\url{http://wise.ssl.berkeley.edu/index.html}}. We adopt the photometry information in $g$, $r$, $z$, $W1$, $W2$ bands from the above datasets to search for the quasar candidates from $425,540,269$ point-like sources in DESI-LIS DR9  catalog\footnote{\url{https://www.legacysurvey.org/dr9/}}). 
    
    \textbf{SDSS eBOSS DR16}

    The Sloan Digital Sky Survey \citep[SDSS, see, e.g. ][for more details]{York2000} is a major multi-spectral imaging and spectroscopic redshift survey and has a long-running history of more than 20 years. The eBOSS \citep[grounded upon SDSS-IV,][]{Blanton2017} is an extended project of BOSS \citep[Baryon Oscillation Spectroscopic Survey, grounded upon SDSS-III,][]{Dawson2013,Eisenstein2011}, which maps the LRGs \citep[luminous red galaxies,][]{Zhou2020, Fortuna2021} and quasars to determine the characteristic scale of BAOs imprinted at the large-scale structure. eBOSS covers a broader range of redshifts than that of BOSS. Based on the Data Release $16$ of eBOSS, \cite{Lyke2020} publishes a catalog containing $750,414$ quasars (DR16Q, hereafter), which is the largest catalog of quasars confirmed spectroscopically. We employ the classification labels from eBOSS DR16 to construct the part of positives in training and testing sets.

    \textbf{SIMBAD}

    SIMBAD is a comprehensive database that collects information on astronomical objects, such as types, fluxes, proper motion, etc., maintained by the Centre de données astronomiques de Strasbourg (CDS). To date, SIMBAD includes $11,953,504$ objects, $\sim$ $50\%$ of them are stars \citep{Paturel2003, Zuckerman2003,Cayrel2004}, and the others are non-stellar objects like AGNs, starburst galaxies, emission-line galaxies \citep{Fu2021}. The types of astronomical objects\footnote{\url{http://simbad.u-strasbg.fr/simbad/sim-display?data=otypes}} in SIMBAD are derived from physical characteristics \citep{Mickaelian2006, Maek2010}, and the astronomical objects with uncertain physical types are marked as `XX\_Candidate' or `Possible\_XX', e.g., `AGN\_Candidate'. In addition, `main\_type' and `other\_types' are given in SIMBAD to deal with the situations in which different studies suggest different types of the same object. We use the non-QSOs data in SIMBAD to build the part of negatives in training and testing sets.
  
    \subsection{Training and Testing Datasets}
    \label{sec:train_and_test}
    The training and testing sets comprise photometry information and labels, where the photometry is from DESI-LIS and WISE, and the labels are from eBOSS DR16 and SIMBAD, respectively. To avoid the problem of overfitting, we extract a subset from the training set for creating a validation set. Notably, the testing set is organized for mimicking the to-be-applied dataset to estimate the classification performance using our method.   
    
    \textbf{Parent Samples}
    
    We first make two parent samples (hereafter, $S1$ and $S2$) to separately prepare the positives and negatives in training and testing sets. For $S1$, we first obtain $745,417$ QSOs\footnote{\url{https://www.legacysurvey.org/dr9/files/\#survey-dr9-region-dr16q-v4-fits}} by combining the photometry in DESI-LIS and labels in DR16Q via cross-matching the catalogs of DESI-LIS and DR16Q. Then, we clean the cross-matched catalog by selecting point-like sources in DESI-LIS (i.e., classified as PSFs) that having all five-bands ($g, r, z, W1, W2$) detections. $S1$ holds $655,017$ QSOs at last. Similarly, $S2$ is acquired by combining photometry in DESI-LIS and the classification labels in SIMBAD, and the cross-match between SIMBAD and DESI-LIS PSFs is executed. Besides, we clean the $1,993,373$ sources obtained through the above procedure according to the `main\_type' in SIMBAD. Details are listed below:
        
    \begin{enumerate}
        \item The sources labeled as quasars are abandoned. 
        
        \item The sources classified by their SEDs \citep[Spectral Energy Distributions, see e.g.][for an introduction]{Richards2006}, region, numbers, time-domain characteristics and the ones with gravitational lensing effect are eliminated. Therefore, the types listed in the first line of Tab.\,\ref{tab:type_exclude} are excluded.
        
        \item The sources that have uncertain physics types are discarded, i.e., we exclude all the sources that have `Candidate', `Possible', `?', or `Unknown' in their `main\_type'.
        
        \item The sources classified as AGN and the types that relate to galaxies are excluded. Therefore, the sources have the labels listed in the second line of Tab.\,\ref{tab:type_exclude} are discarded. Although we only use point-like sources in DESI, some galaxies are still involved in DESI-LIS PSFs because extremely compact galaxies and the high-density regions in large galaxies might be classified as point sources.
    
        \item The `LINER', `Blazar', `Seyfert' and `BLLac' are cleaned since they mimic the color of the quasars \citep{Peters2015}. 
        
    \end{enumerate}

    After these operations, we further take care of the information given in `other\_types'. We remove the sources that have the labels listed in the third line of Tab.\,\ref{tab:type_exclude}. At the end, there are $1,363,030$ non-QSOs left from $1,993,373$ sources. Note that $99.85\%$ of non-QSOs are stars, and the rests have non-stellar features such as `HII (ionized) region' ($0.11\%$) and `Emission Object' ($0.02\%$). Details of the catalog of negatives are available online \footnote{\url{https://github.com/EigenHermit/he-li2021/blob/main/s2\_types.csv}}.

    \begin{table*}[]
        \centering
        \begin{tabular}{c|c}
        \hline
                           & SIMBAD Type                                                  \\ \hline
        \multirow{2}{*}{1} & Radio/Region/Gravitation/lensedimage/Lensed/GravLens         \\
                           & Void/Transient/Maser/IR/Red/Blue/UV/X/gamma/multiple\_object \\ \hline
        \multirow{3}{*}{2} & SuperClG/ClG/GroupG/Compact\_Gr\_G/PairG/IG/OpCl             \\
                           & GinPair/LISB\_G//HII\_G/GinGroup/PartofG/EmG/GinCl/          \\
                           & Galaxy/AGN                                                   \\ \hline
        \multirow{2}{*}{3} & quasar/Q?/AGN/Galaxy/G/Gravitation/grv/Lev/LIS?/Le?/LI?/gLe? \\
                           & gLIS/GWE/reg/vid/SCG/CIG/CrG/CGG/PaG/IG                      \\ \hline
        \end{tabular}
        \caption{The types of objects that we abandon when build the training and testing set. See more details at Sec.\,\ref{subsec:AvlblObsrvtns}}.
        \label{tab:type_exclude}
    \end{table*}

    \textbf{Testing Set}
        
\begin{figure}
        \centering
        \includegraphics[scale=0.35]{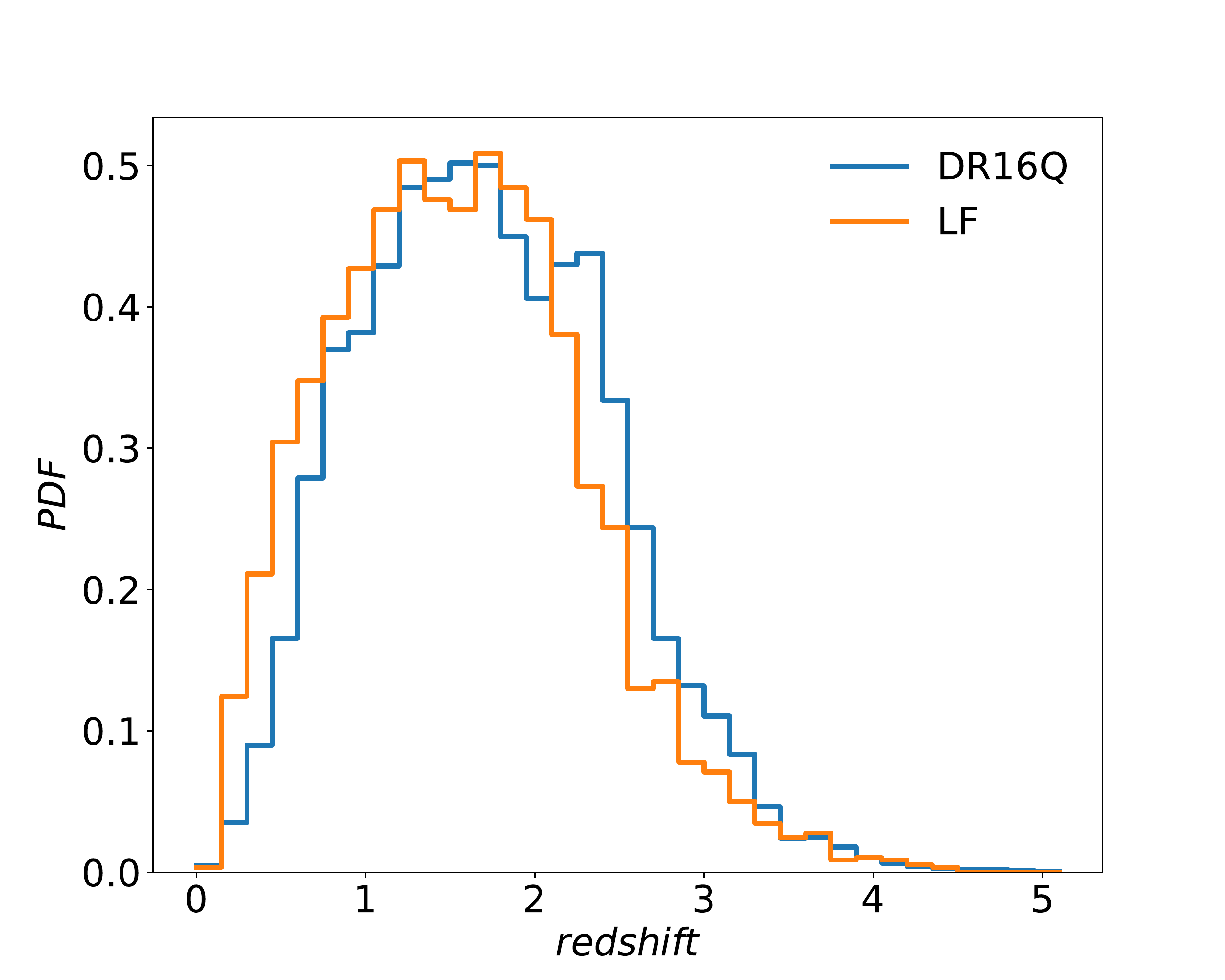}
        \caption{The comparison of redshfit distributions. The orange one is calculated by \eqref{eq:LF}. The blue one is from DR16Q.}
        \label{fig:z_chk}
    \end{figure}

    The testing set are constructed by selecting objects in $S1$ and $S2$ according to the distributions of magnitudes of QSOs in DESI-LIS modeled by a typical luminosity function (LF, hereafter) and a SED  \citep{Bianchini2019} since the distributions of QSO and non-QSO in testing set should be similar to the corresponding ones of DESI-LIS if we plan to evaluate the classification performance with the testing set. The LF is double power-law as is modeled in OM10:
    \begin{equation}
        \label{eq:LF}
            \frac{d\Phi_{QSO}}{dM}=\frac{\Phi_*}{10^{0.4(\alpha+1)(M-M_*)}+10^{0.4(\beta+1)(M-M_*)}}\,\, ,
        \end{equation}
where $M$ stands for the absolute $i$-band magnitude of quasars. $M_*$ indicate the change of LF with redshift, which is given by
        \begin{equation}
            M_*=-20.90+5{\rm log}\ h-2.5{\rm log}\ f(z)
        \end{equation}
        \begin{equation}
            f(z)=\frac{e^{\zeta z}(1+e^{\xi z_*})}{(\sqrt{e^{\xi z}}+\sqrt{e^{\xi z_*}})^2}
    \end{equation}
        The parameter settings ($\zeta, \ \xi, \ \alpha, \ \beta$) are also taken from OM10. On the other hand, the number of quasars in a redshift bin $(z_0, z_1)$ is 
    \begin{equation}
        \label{equ:z}
            N(z_0, z_1) = V \int_{M_2}^{M_1}\frac{d\Phi_{QSO}}{dM} dM \,\,,
    \end{equation}
    where $V$ is the comoving volume in $(z_0, z_1)$, $\frac{d\Phi_{QSO}}{dM}$ is given by Equ.\,\ref{eq:LF}. $M_1$ and $M_2$ are the upper and lower boundary of $M$. The comparison of the redshift distributions given by LF and by observations is show in Fig.\,\ref{fig:z_chk} and they are in a good agreement. 

\begin{figure}
        \centering
        \includegraphics[scale=0.35]{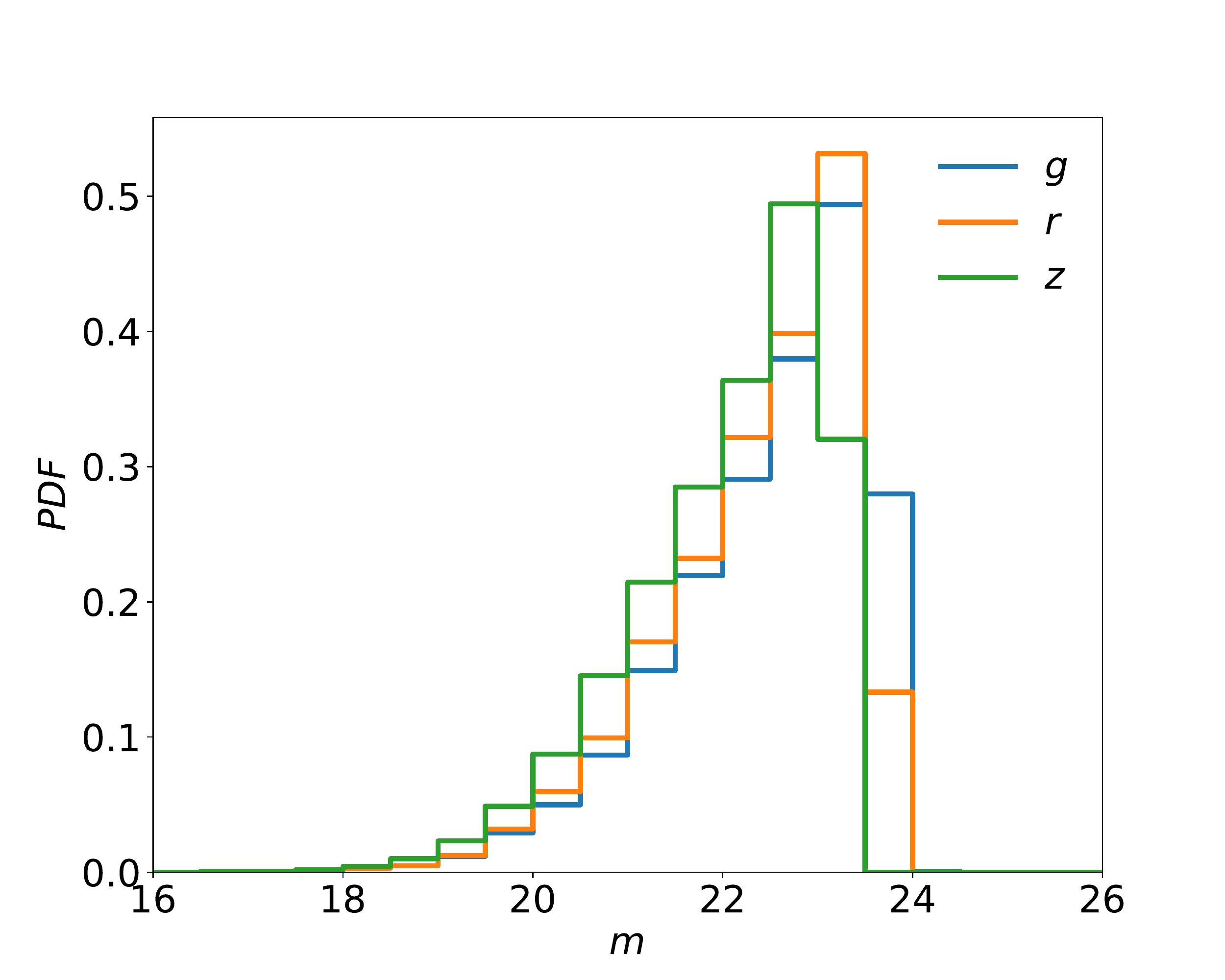}
        \includegraphics[scale=0.35]{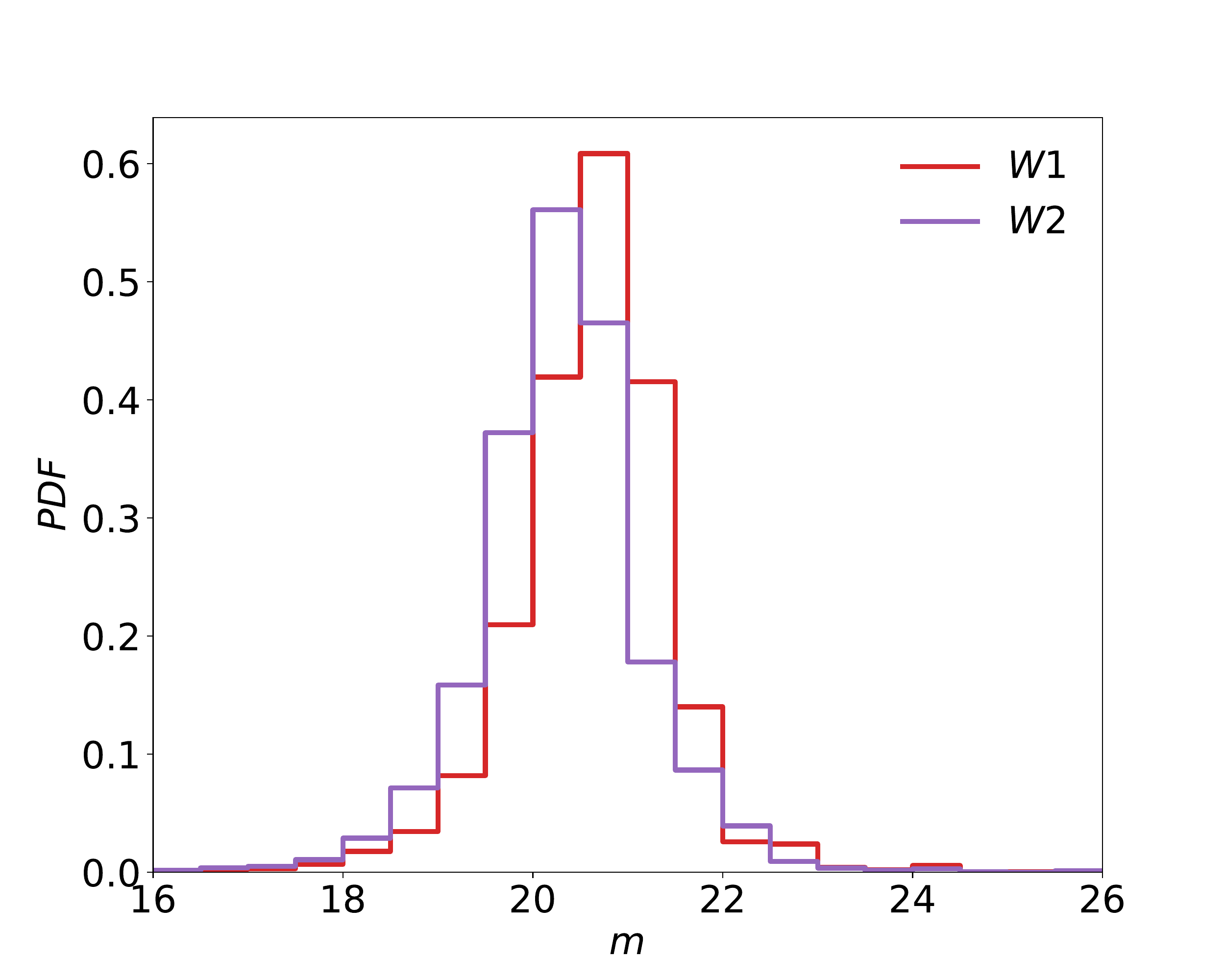}
        \caption{The distributions of $g,r,z,W1,W2$ of the quasars in the testing set.}
        \label{fig:tstqso}
    \end{figure}

    We further predict the magnitude distributions of  $g$, $r$, $z$-bands using a typical SED \citep{Bianchini2019} based on the $i$-band magnitude distribution of QSOs acquired above. According to the predicted distributions of $g$, $r$, $z$-bands, the $W1, W2$ is directly taken from $3,944$ quasars in $S1$. The $W1, W2$ cannot be calculated merely based on SED because the host galaxies of QSOs contribute to the total fluxes considerably in $W1, W2$ bands \citep{Li2021}. Thus, the $10,000$ mock quasars form the positives in the testing set, and the distributions are shown in Fig\,\ref{fig:tstqso}.  
                
\begin{figure}
        \centering
        \includegraphics[scale=0.35]{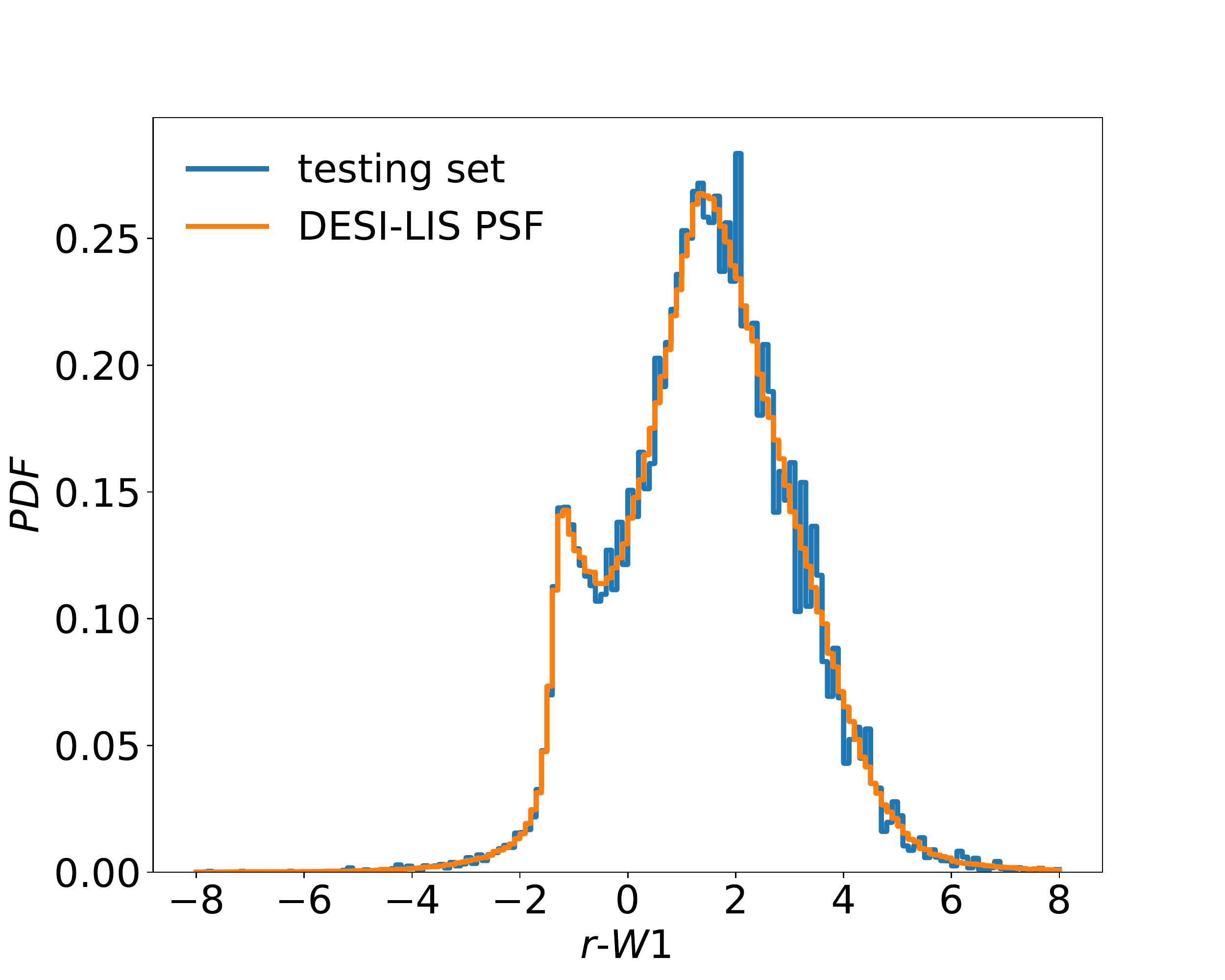}
        \includegraphics[scale=0.35]{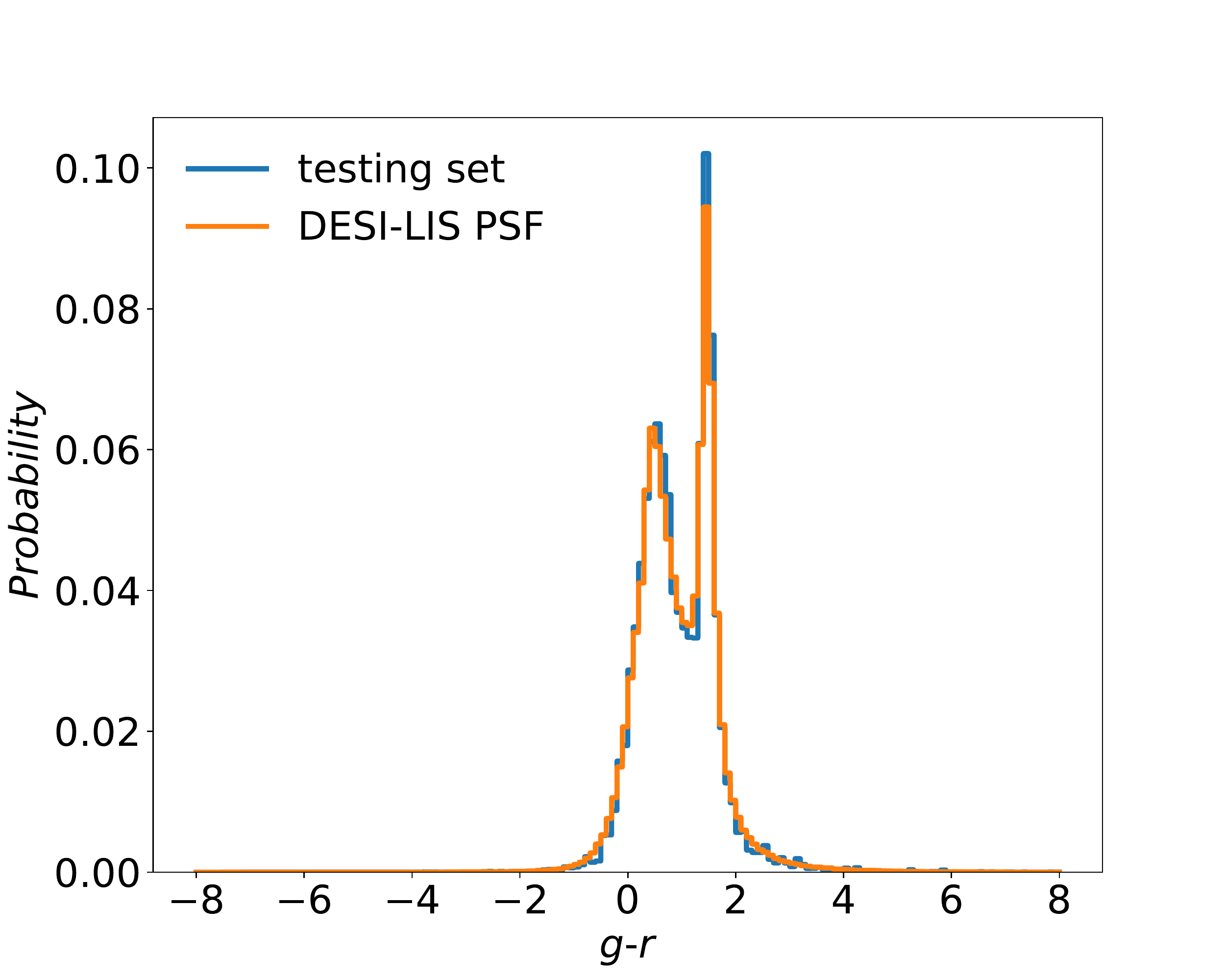}
        \caption{The comparisons between the testing set and DESI-LIS PSFs. The upper is in color space while the lower in magnitude space.}
        \label{fig:mag_clr_compare}
    \end{figure}

    Moreover, we select the non-QSOs from $S2$ to construct the negative part of the testing set. The criterion is that: we find such non-QSOs to make the overall distributions (testing QSOs + non-QSOs) similar to the overall distributions of DESI-LIS PSFs. The testing set contains $990,000$ non-quasars ($135,858$ individual ones). The comparisons in magnitude and color spaces between DESI-LIS PSFs and the testing set are shown in Fig.\,\ref{fig:mag_clr_compare}. In the space of all $15$-d colors, the twos show a good agreement and are particularly good in the color space (the mean and stranded deviation differences are less than $1\%$). In magnitude space, however, the testing set is slightly brighter than DESI-LIS PSFs due to the selection effect of SIMBAD.

    Above all, the testing set contains $100,000$ sources, 1\% \citep{Page2001} of them are quasars, built upon $3,944$ quasars in $S1$ and $135,858$ non-quasars in $S2$. 

    \textbf{Training Set}
        
\begin{figure}
        \centering
        \includegraphics[scale=0.35]{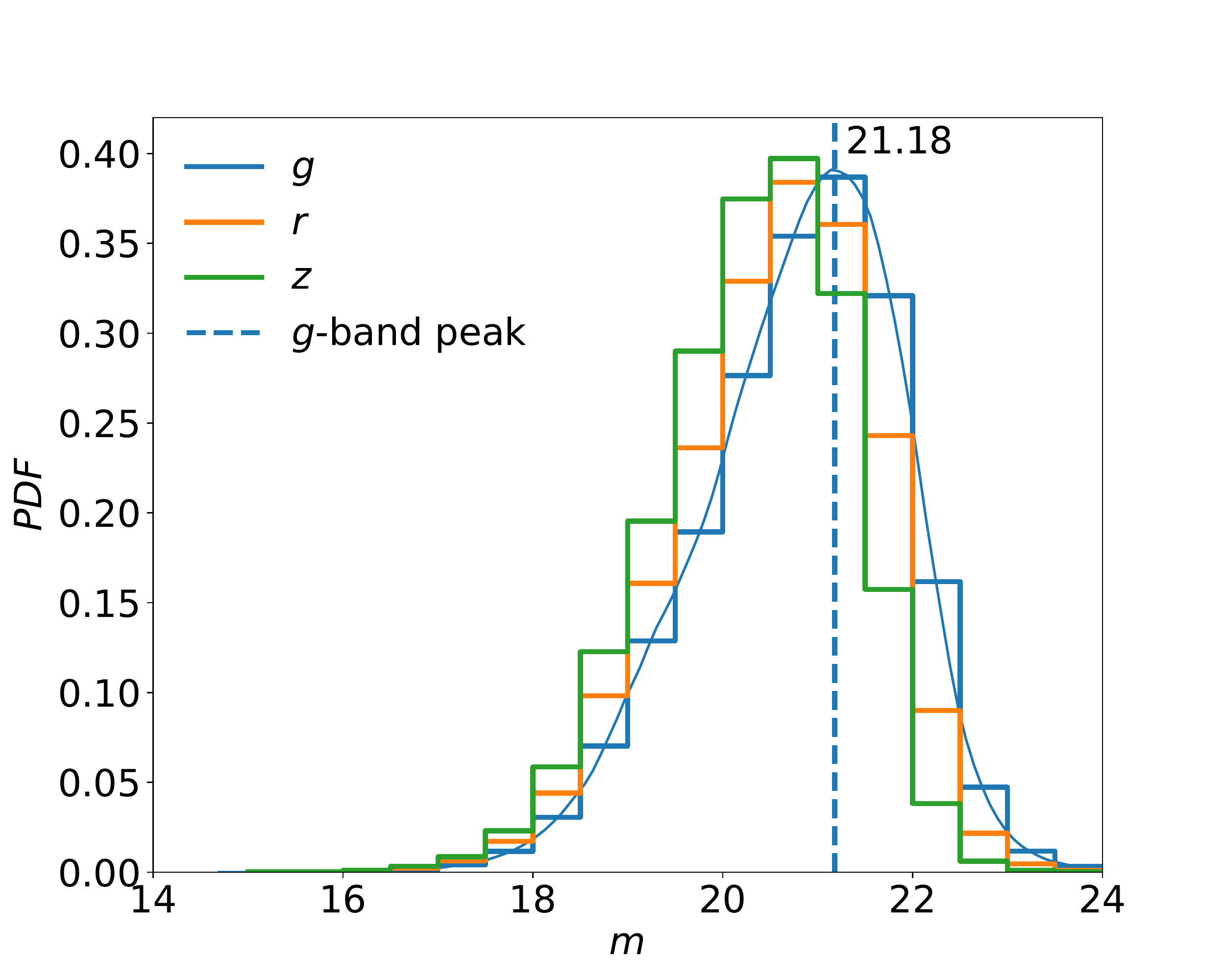}
        \includegraphics[scale=0.35]{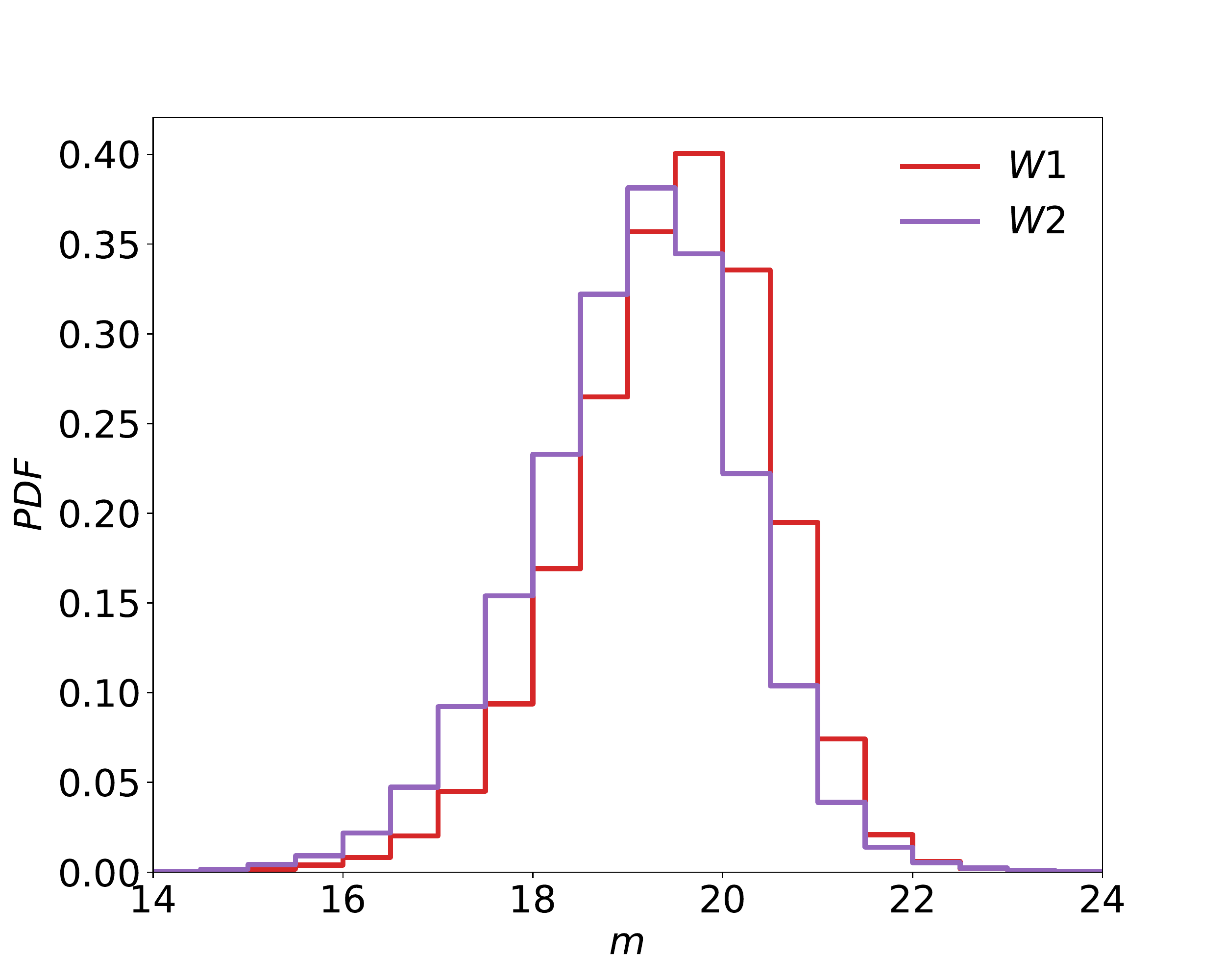}
        \caption{The distributions of $g,r,z,W1,W2$ of the quasars in the training set. The $g$-band peak magnitude is shown by the vertical dashed line, which is determined by a kernel density estimate plot (blue solid curve) that gotten by \textit{\textbf{kdeplot}} in \textit{\textbf{seaborn}} package with the default Gaussian kernel and $binwidth=0.5$.}
        \label{fig:traing_set_qso}
    \end{figure}

    Except for the sources in the testing set, the rests in $S1$ and $S2$ comprise the training set. Explicitly, there are $651,073$  positives and $1,227,172$  negatives in the training set, and the distributions of quasars in the training set are shown in Fig.\,\ref{fig:traing_set_qso}. 
 
\section{Methodology} 
\label{sec:mthd}
    The primary approach for identifying QSO candidates is constructed upon Random Forest in this work. The evaluation metrics for the outcomes include completeness, accuracy, and area under the receiver operating characteristic (ROC) curve (AUC). Also, we create a baseline for the identification of QSO candidates using the traditional color-cut selection method for cross-validation.

    \subsection{Random Forest}
    \label{sec:rf}
    Random Forest is a mature machine learning algorithm and has been widely employed in astronomy. RF was firstly proposed and named `random decision forests' by \cite{Ho1995}, then improved and renamed `random forests' by  \cite{Breiman2001}. The basic workflow of RF is that:  1)Randomly segments the input data; 2) trains a group of decision tree models \citep{Dobra2018} with the segmented data separately; 3) gives the final judgments by combining the outputs of all decision trees. \cite{Breiman2001b} has suggested that RF compare favorably to AdaBoost \citep{Freund1996}  but it is more robust on missing and unbalanced data, and performs well on multi-dimension data.

    In particular, when dealing with the classification of star-galaxy-quasar with photoemtry, \cite{Bai2019} presents the superiority of RF over K-Nearest Neighbour \citep{Altman1992} and Support Vector Machine \citep{Cristianini2008} by implementing a comprehensively comparative investigation, which inspires us to choose RF for our purpose.
    
    In this study, we construct our classifier based on the RF module in \textit{\textbf{Scikit-learn}} package \footnote{\url{https://scikit-learn.org/stable/}} \citep{Pedregosa2011}. The parameters of RF model are tweaked to achieve the best completeness and purity (described at Sec\,\ref{sec:metrics}) that evaluated by validation set, explicitly, $max\_depth=20$, $n\_estimators=200$, $oob\_score=True$, and $random\_state=0$. 
    
    \subsection{Color-cut Selection}
    \label{sec:cc}
    The color-cut selection \citep[in $g-z$ versus $grz-W$ space, following][]{Yeche2020} is performed to validate the candidates that selected by Random Forest model, which is a widely used method for selecting quasars with photometry data \citep[see e.g.][for the applications]{Warren1991,Croom2004,Richards2005,Morganson2014} because the magnitude of quasars in the UV band is brighter than normal stars and galaxies \citep{Elvis1994} comparing to the stars that have similar magnitudes in optical bands. QSOs are also roughly two magnitudes brighter in the near infrared bands across a wide redshift range \citep{Peters2015}. Above all, we slightly update the strategies of color-cut selection procedure in \cite{Yeche2020}, and create a QSOs candidate catalog for cross-validating with the results given by RF model. Details of selection criteria are listed in Tab.\,\ref{tab:c-c_selection}, and the definitions are:
    \begin{equation}
    \begin{split}
        flux(grz) &= \frac{flux(g)+0.8\times flux(r)+0.5\times flux(z)}{2.3} \\
        flux(W) &= 0.75\times flux(W1)+0.25\times flux(W2)
    \end{split}
    \end{equation}
     The first four criteria in Tab.\,\ref{tab:c-c_selection} are directly taken from \cite{Yeche2020} while the last one is designed to further limit the samples by taking advantage of the aforementioned infrared excess. We note that we use depth limit \citep[$g=22.7$, similar with ][]{Yeche2020} in this color-cut selection.
       
    \begin{table}[]
    \centering
    \begin{tabular}{c|c}
    \hline
      & Conditions                       \\ \hline
    1 & $g$-$r$\textgreater{}1.3             \\ \hline
    2 & $-0.4$\textless{}$r$-$z$\textless{}$1.1$ \\ \hline
    3 & $r$\textgreater{}$17.5$             \\ \hline
    4 & $grz$\textgreater{}$17.0$            \\ \hline
    5 & $-1$\textless{}$grz$-$W$\textless{}$4$   \\ \hline
    \end{tabular}
    \caption{The selection conditions that used in color-cut selection (detailed at Sec.\,\ref{sec:cc}.)}
    \label{tab:c-c_selection}
    \end{table}
    
    \subsection{Evaluation Metrics}
    \label{sec:metrics}
    Completeness, accuracy, and AUC are quintessential classification metrics, which measure the performance of classification models from various angles. The three quantities can be calculated by combining True Positives (TP), False Positives (FP), True Negatives (TN), and False Negatives (FN) in different forms. Specifically, completeness is given by
    \begin{equation}
        completeness = \frac{TP}{TP+FN}, 
    \end{equation}
    meaning the percentage of quasars in the testing set that can be correctly picked out; accuracy is given by
    \begin{equation}
        accuracy = \frac{TP+TN}{TP+TN+FP+FN}
    \end{equation}
    standing for how well the identification of quasars/non-quasars is; purity is used to assess how much non-quasar contamination in the quasar candidate catalogs, which is defined as
    \begin{equation}
        purity = \frac{TP}{TP+FP}\,\, ;
    \end{equation} 
    AUC is the area under the ROC \citep{Fawcett2006}, showing the performance of the classification at all classification thresholds by plotting False Positive Rate (FPR) versus True Positive Rate (TPR), defined as
    \begin{align}
        FPR = \frac{FP}{FP+TN}\,\, ,
        TPR = \frac{TP}{TP+FN}\,\, .
    \end{align} 
    AUC represents the overall performance of identification results. 
  
    \section{Results}
    \label{sec:cnclsn}
    In results of applying our classifier trained by the training set given in Sec.\,\ref{sec:train_and_test} to DESI-LIS point-like sources (Sec\,\ref{subsec:AvlblObsrvtns}), we accomplish a catalog of quasar candidates, detailed in Sec.\,\ref{sec:catalog}. Then, we cross-validate the catalog with the results obtained via color-cut selection method (see Sec.\,\ref{sec:cc}), and the details are shown in Sec.\,\ref{sec:Cross-validation}.

    \subsection{Quasar Candidates Catalogs}
    \label{sec:catalog}
    
\begin{figure}
        \centering
        \includegraphics[scale=0.35]{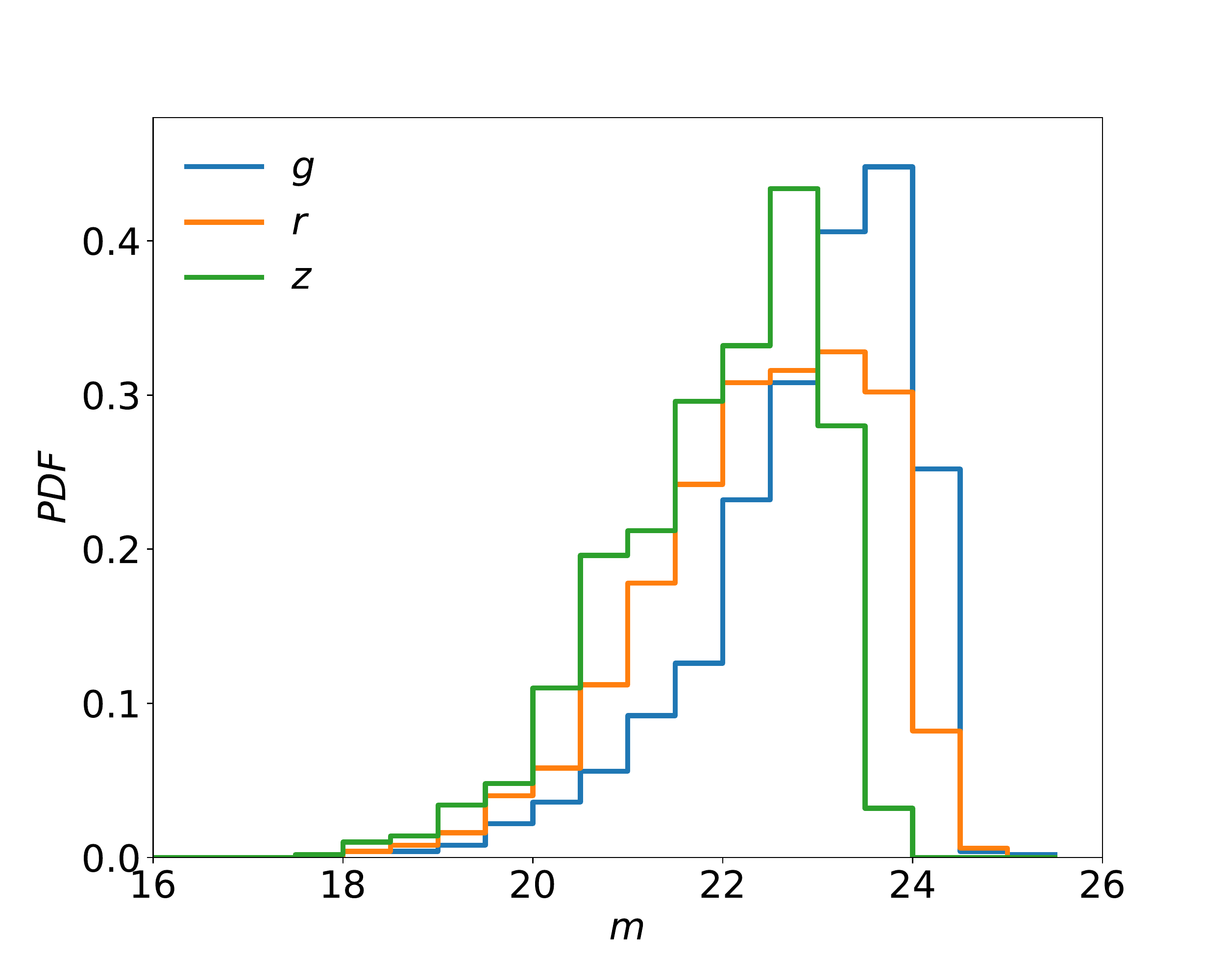}
        \includegraphics[scale=0.35]{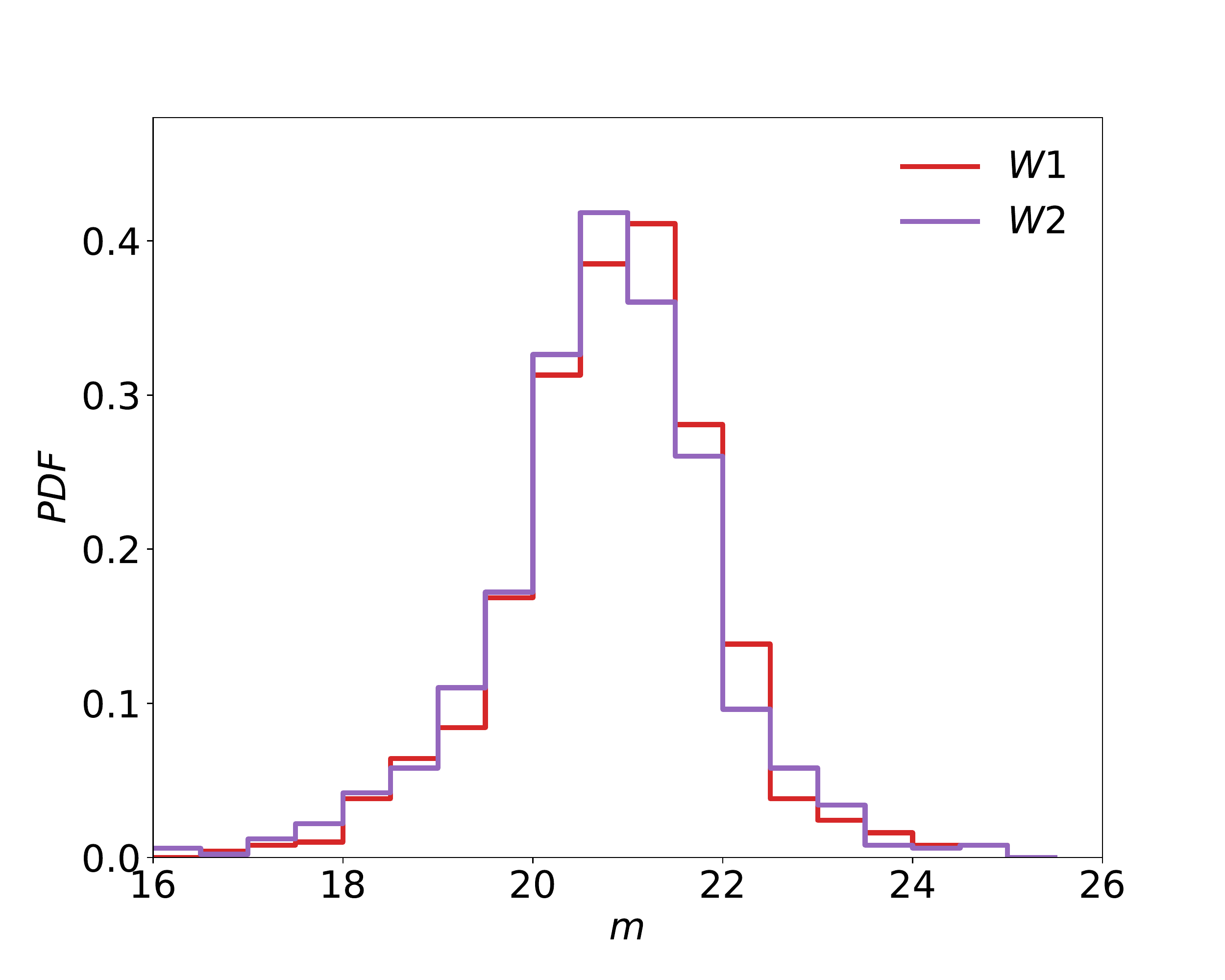}
        \caption{The distributions of $g,r,z,W1,W2$ of all quasars candidates.}
        \label{fig:dist_qcc}
    \end{figure}

\begin{figure}
        \centering
        \includegraphics[scale=0.35]{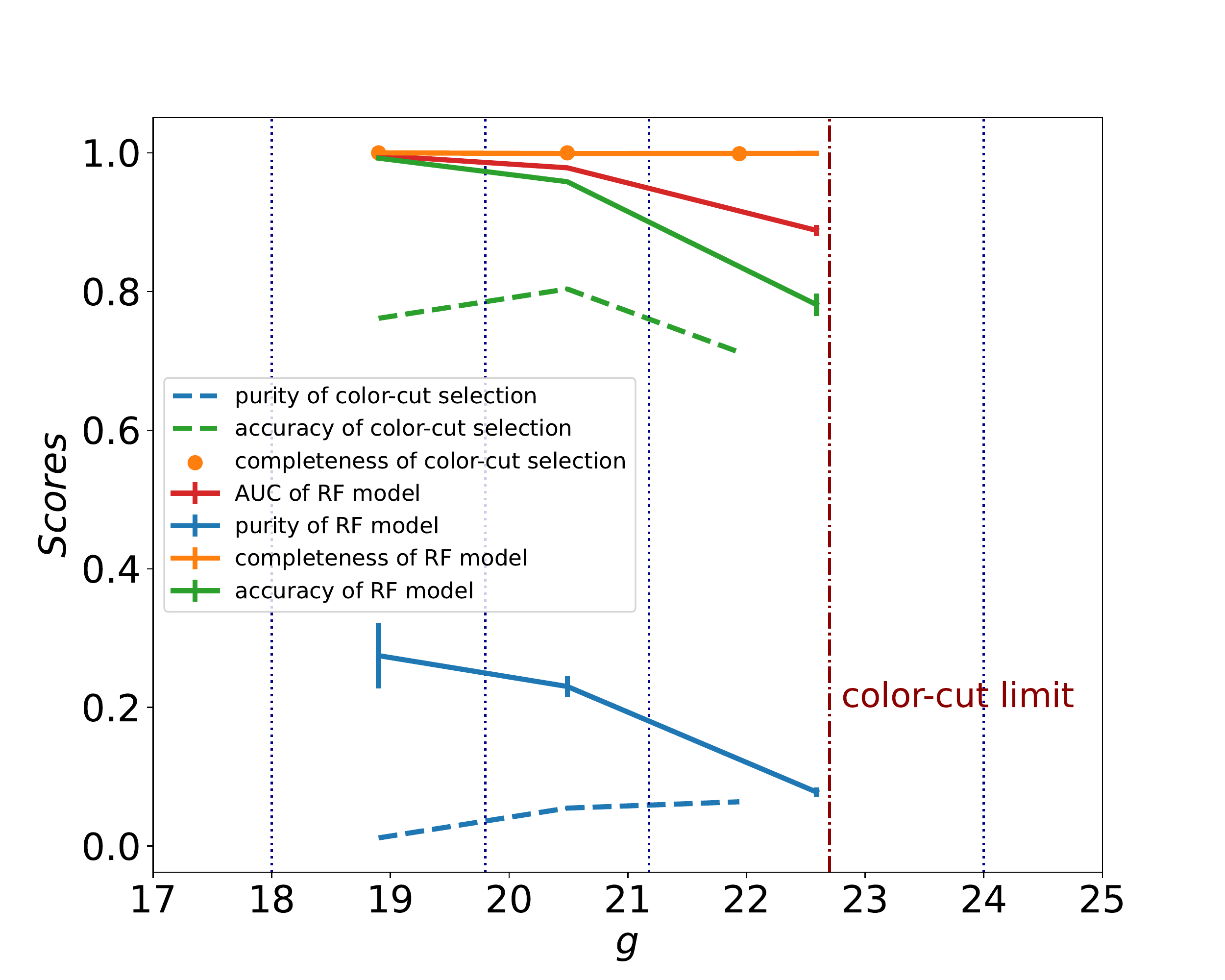}
        \caption{The results when the RF model (solid lines) and color-cut selection (dashed lines) are applied to the testing set. The testing set is divide in to three subsets according to their $g$-band magnitudes. The details are given in Sec.\,\ref{sec:catalog}. When testing the color-cut selection, a limit magnitude is used (detailed at Sec.\,\ref{sec:cc}) and shown with red dot-dashed line.}
        \label{fig:magbin_tst}
    \end{figure}

    We acquire $24,440,816$ quasar candidates, and the magnitude distributions is shown in Fig\,\ref{fig:dist_qcc}. To evaluate the completeness, purity, accuracy, and AUC of the classification, our RF model is applied to the testing set, and the result is shown in Fig.\,\ref{fig:magbin_tst}. The completeness remains stable while the other metrics drop significantly at the faint end. The three bins of the testing set are divided by $g$-band magnitudes given below, and each bin contains similar amounts of unique sources:
    \begin{center}
        bin1: $18<g<19.8$\\
        bin2: $19.8<g<21.18$\\
        bin3: $21.18<g<24$\,\,.
    \end{center}
    We split the sample into Grade-A and Grade-B with a $g$ band magnitude of $21.18$ because the purities of classifying the objects below and above the magnitude are significantly different. Grade-A and Grade-B contain $1,953,932$ and $22,486,884$ candidates separately.

    The above classification is constructed by defining a threshold of the probability given by our classifier, where $P_{\rm th} = 0.5$. Correspondingly, Grade-A has the purity $\sim 0.3$ and accuracy $\sim 0.99$; Grade-B has the purity $\sim 0.15$ and accuracy $\sim 0.90$. Notably, the completeness remains high in all magnitude bins (see the orange line in Fig.\,\ref{fig:magbin_tst}). The errorbars represent the uncertainties by bootstrapping the elements in each bin \citep{Efron1982}. Expressly, larger errorbars of purity in brighter bins are due to the fewer quasars; larger errorbars of accuracy in fainter bins are due to the decreasing capacity of the RF model.

\begin{figure}
        \centering
        \includegraphics[scale=0.35]{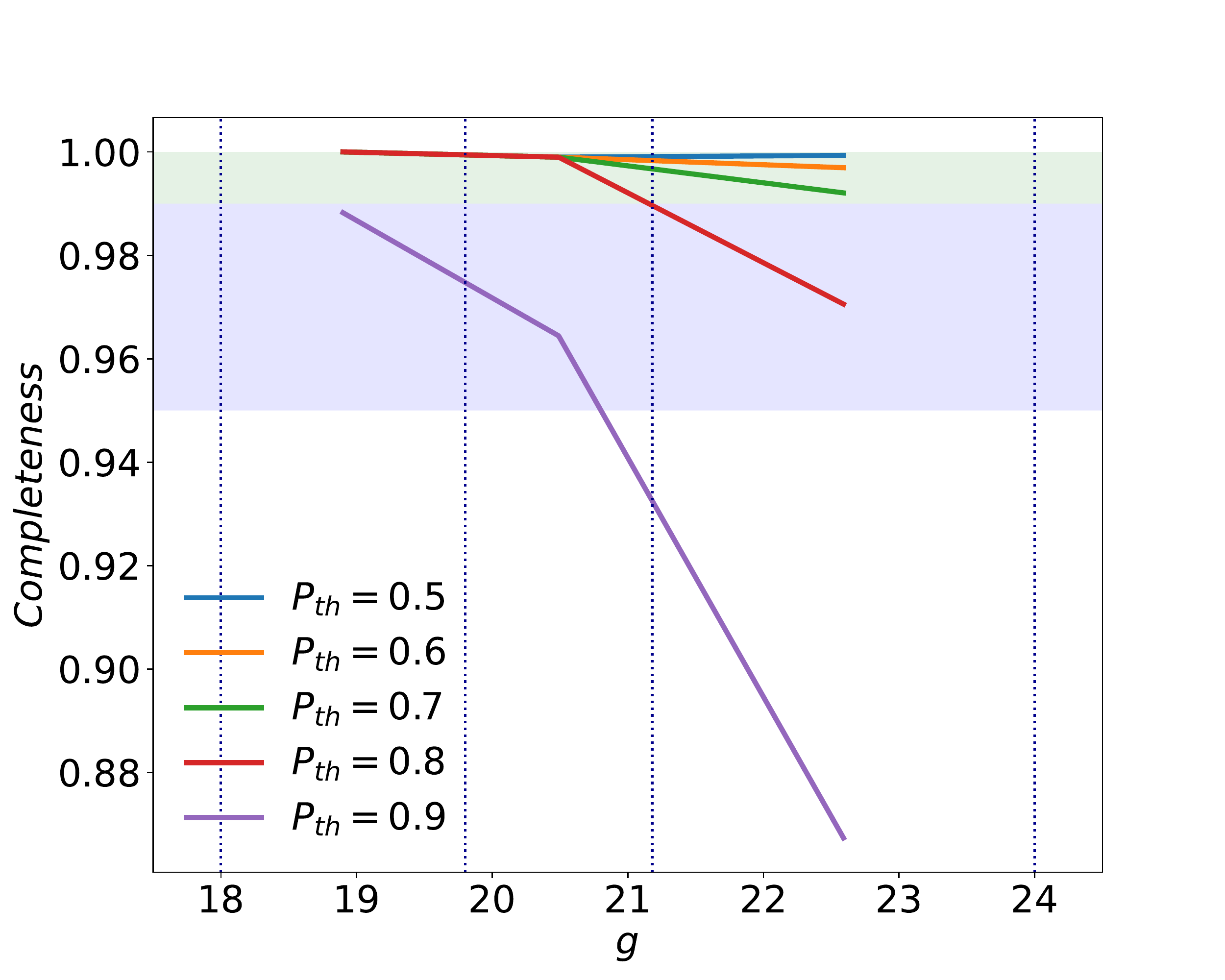}
        \caption{The variety of completeness when different $P_{\rm th}$ are applied. The way of splitting testing set is the same with the ones for Fig.\,\ref{fig:magbin_tst}. The area where the completeness higher than $0.99$ is filled by green, while the ones higher than $0.95$ is filled by purple.}
        \label{fig:comp_tst}
    \end{figure}

    To further investigate the effects of the thresholds of the probability of being a QSO on the performance of the classification, we test $0.5,0.6,0.7,0.8,0.9$ as $P_{\rm th}$, and the result is shown in Fig.\,\ref{fig:comp_tst}. In all cases, the completeness is higher than $0.85$, and the completeness is even higher than $0.99$ (see the green filled region) when $P_{\rm th} = \{0.5, 0.6, 0.7\}$, but it decreases to $0.95$ (see the purple filled region) when $P_{\rm th} = 0.8$. Therefore, $P_{\rm th} = 0.5$ is for general purpose, and it gives high completeness but low purity. However, one can change $P_{\rm th}$ for a specific combination of completeness and purity according to the tendency shown in Fig.\,\ref{fig:comp_tst}.

\begin{figure*}
        \centering
        \centerline{\textbf{Grade-A}}
        \includegraphics[scale=0.4]{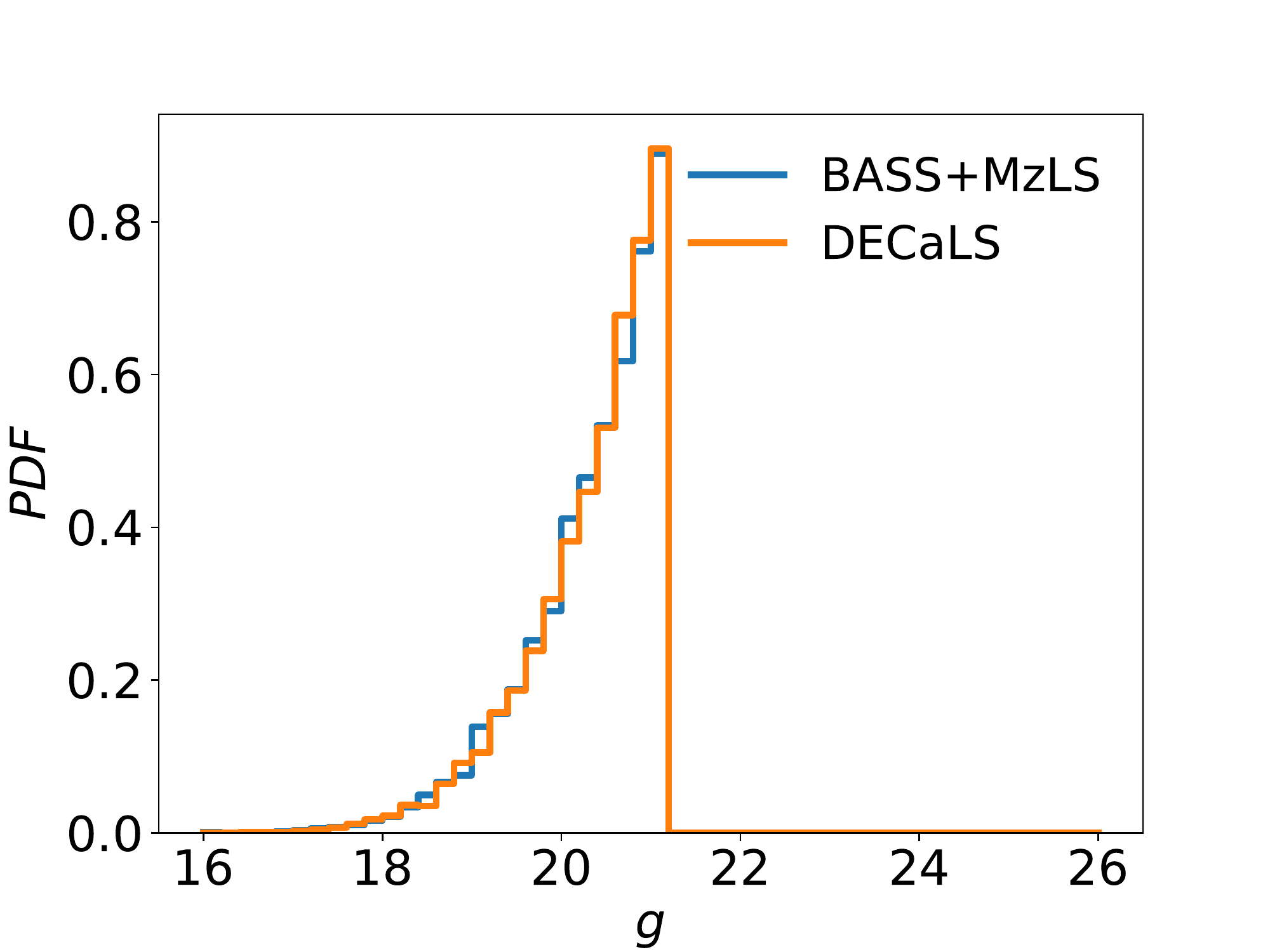}
        \includegraphics[scale=0.4]{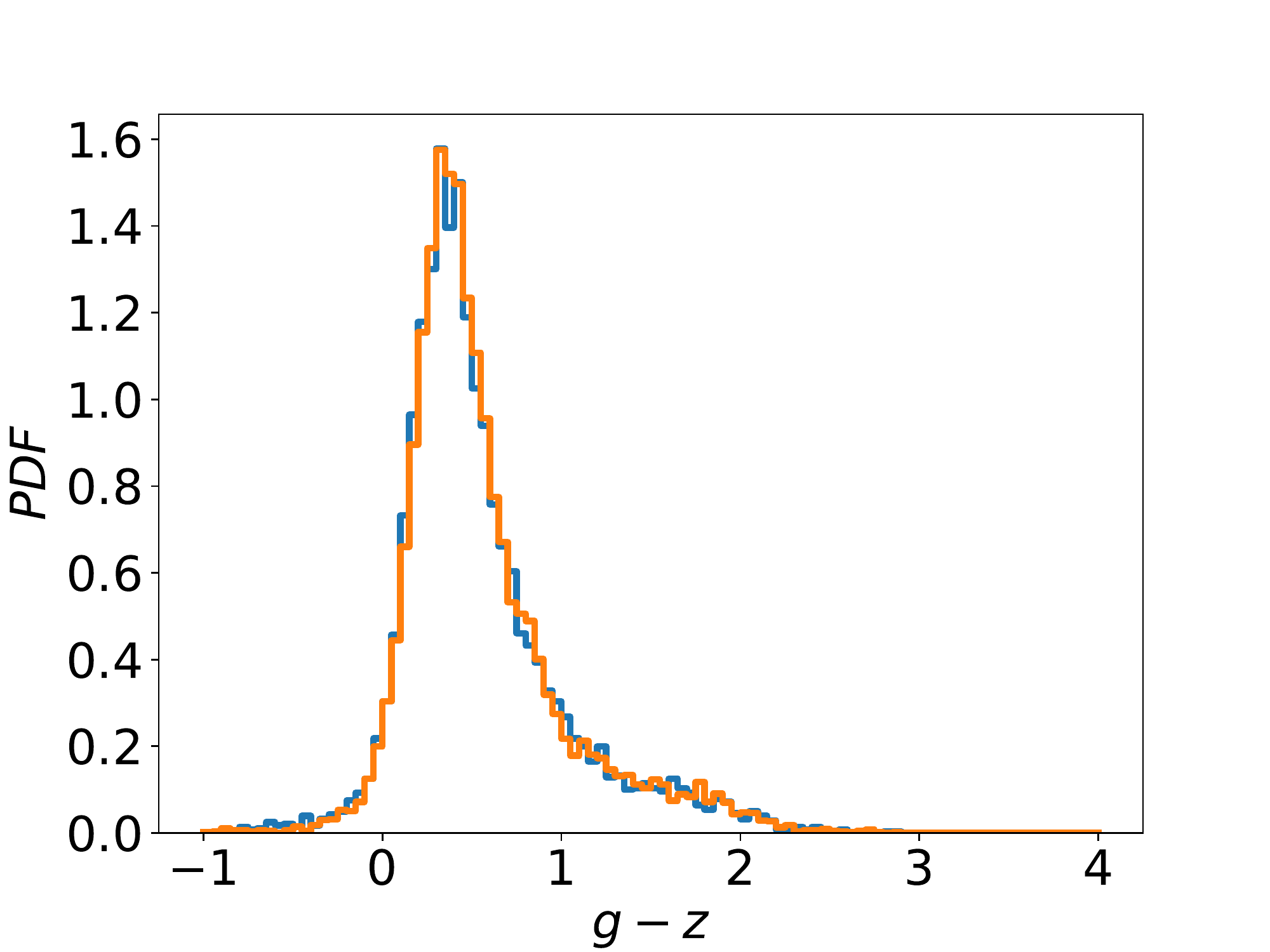}
        \centerline{\textbf{Grade-B}}
        \includegraphics[scale=0.4]{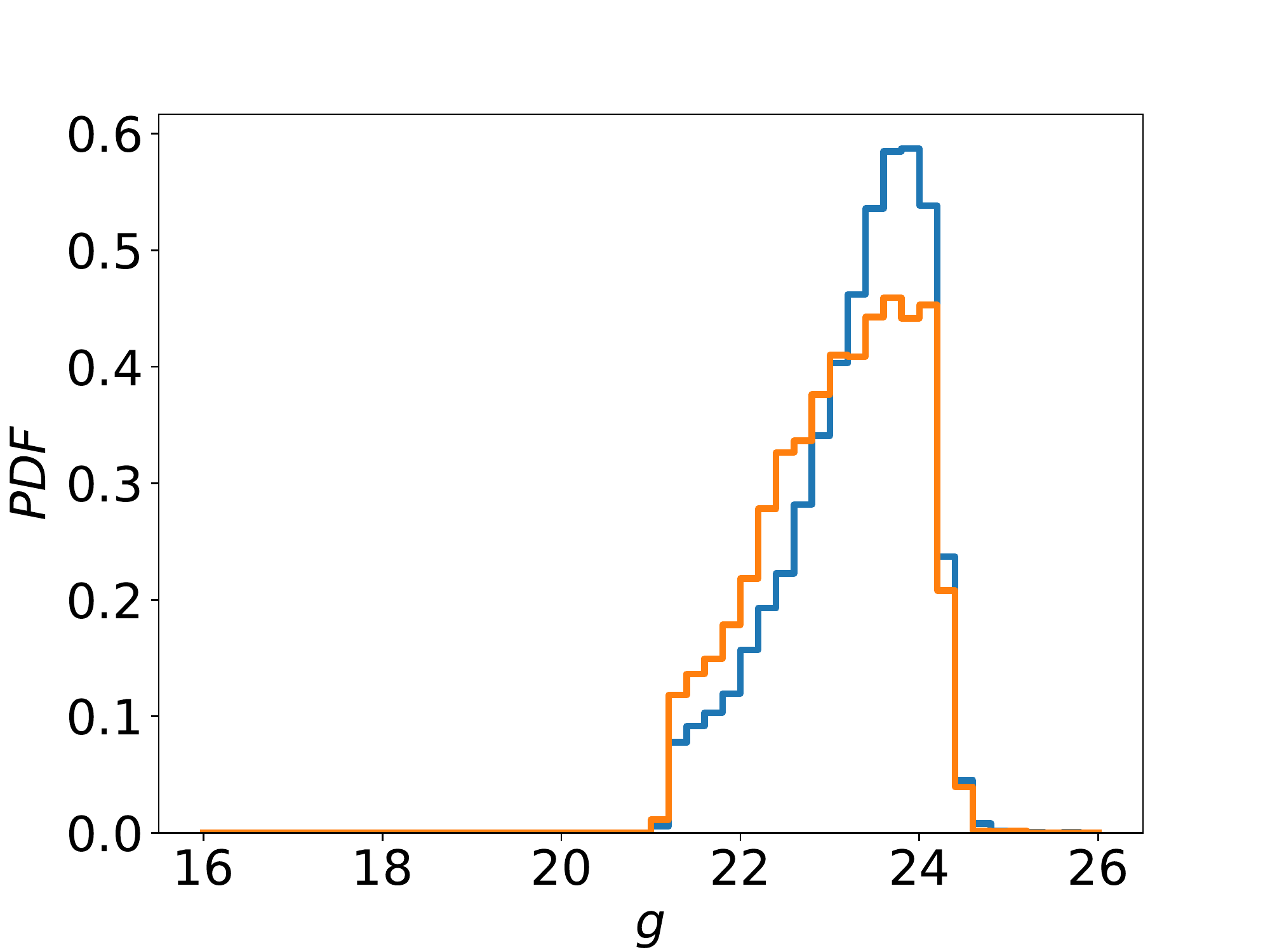}
        \includegraphics[scale=0.4]{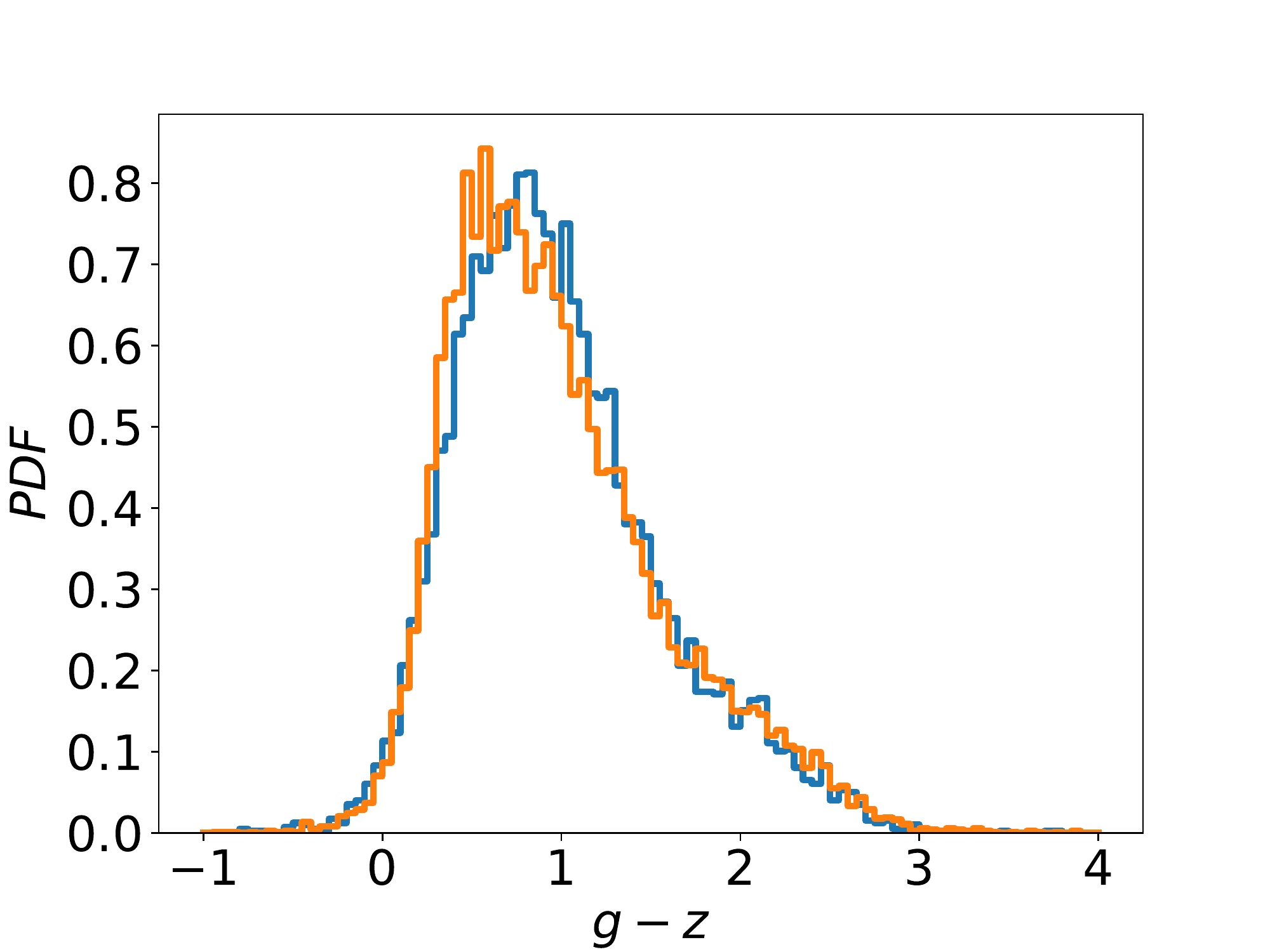}
        \caption{The $g$ and $g-z$ distributions of Grade-A (first line) and Grade-B (second line) candidates that been observed in BASS + MzLS and DECaLS footprints. }
        \label{fig:decals_bass}
    \end{figure*}

\begin{figure}
        \centering
        \includegraphics[scale=0.35]{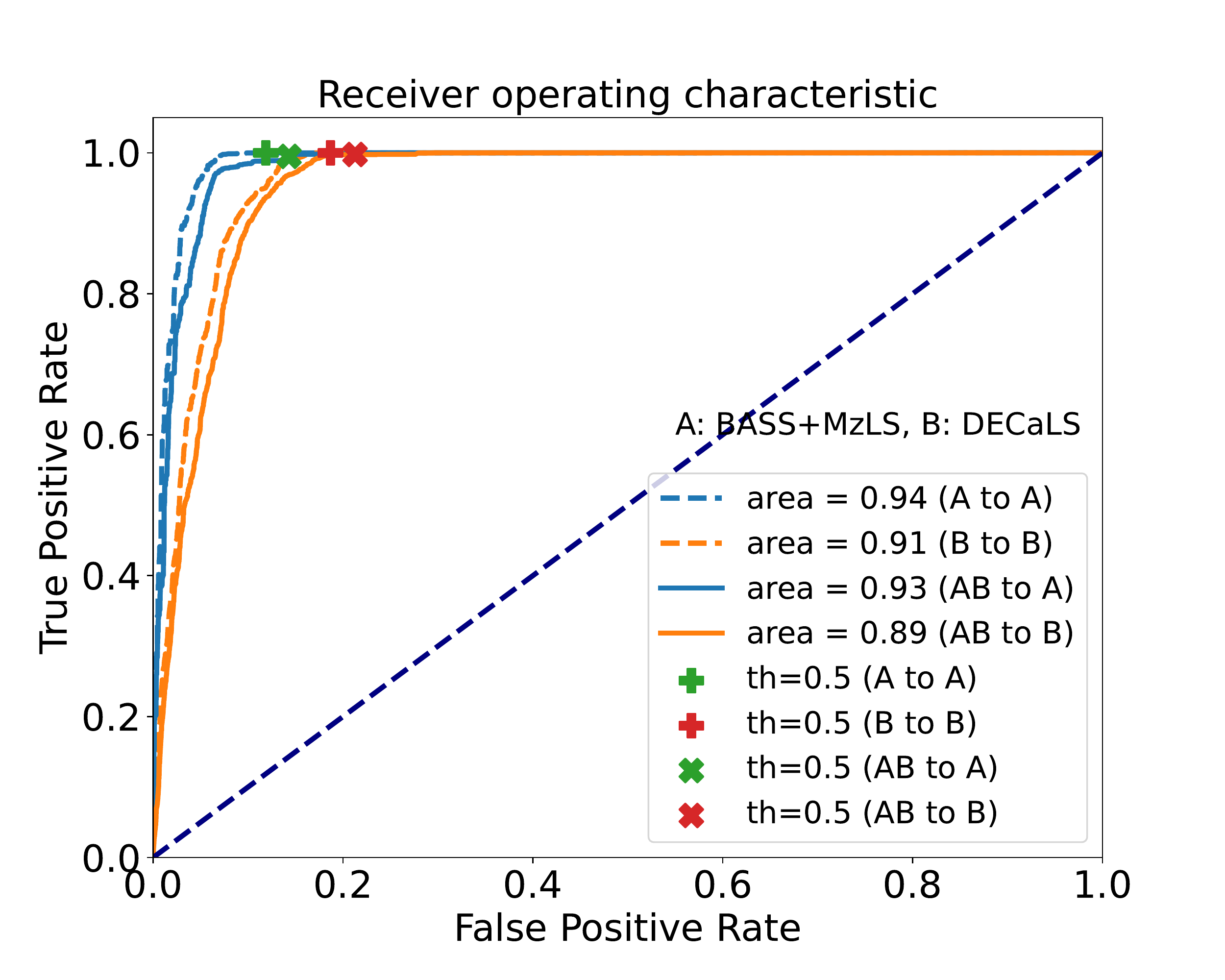}
        \caption{The ROC curves of the RF model that is trained in four different cases. The four scatters explicitly indicate the FPR and TPR when $P_{\rm th}=0.5$.}
        \label{fig:roc}
    \end{figure}

    Considering the various setups of DECaLS and BASS+MzLS, the corresponding QSO candidates in their footprints have different statistical properties. As displayed in Fig.\ref{fig:decals_bass}, the $g$-band magnitudes distribution and $g-z$ distribution are identical in DECaLS and BASS + MzLS footprints for Grade A candidates due to their higher confidence. But for Grade B candidates, the most significant difference between the magnitude distributions of the candidates in DECaLS and BASS + MzLS footprints is in the $g$-band because the most notable difference is in the efficiency of the $g$-filter of DECaLS and BASS. The Grade B candidates in DECaLS are shallower than those in BASS + MzLS in the $g$-band and are slightly bluer than those in BASS + MzLS when choosing $g-z$ as the color indicator. To quantify the influences of the above differences on the identification of quasar candidates, we explore the performance of the classification when the RF models are individually trained by DECaLS (or BASS+MzLS). As shown in Fig.\,\ref{fig:roc}, the overall performances (represented by the area under the ROC, a.k.a AUC) with the new training strategy are slightly better than the earlier one which ignores the different setups between DECaLS and BASS + MzLS. However, the differences in completeness (i.e., TPR) are $\sim 1$ \textperthousand $\ $when we choose 0.5 (the adopted value in this work) as the classification threshold, while FPRs and Purity increase $10 \sim 20\%$ (see cross-marks and plus-marks in Fig.\,\ref{fig:roc}, and Tab.\,\ref{tab:whenth0.5} for explicit values). For others' convenience, we attach catalogs of quasar candidates obtained by adopting the latter training and classifying strategy as complementary to the earlier results, which can be found in the same repository \footnote{\url{https://github.com/EigenHermit/he-li2021}}.

    
    
    \subsection{Cross-validation}
    \label{sec:Cross-validation}
\begin{figure}
        \centering
        \includegraphics[scale=0.35]{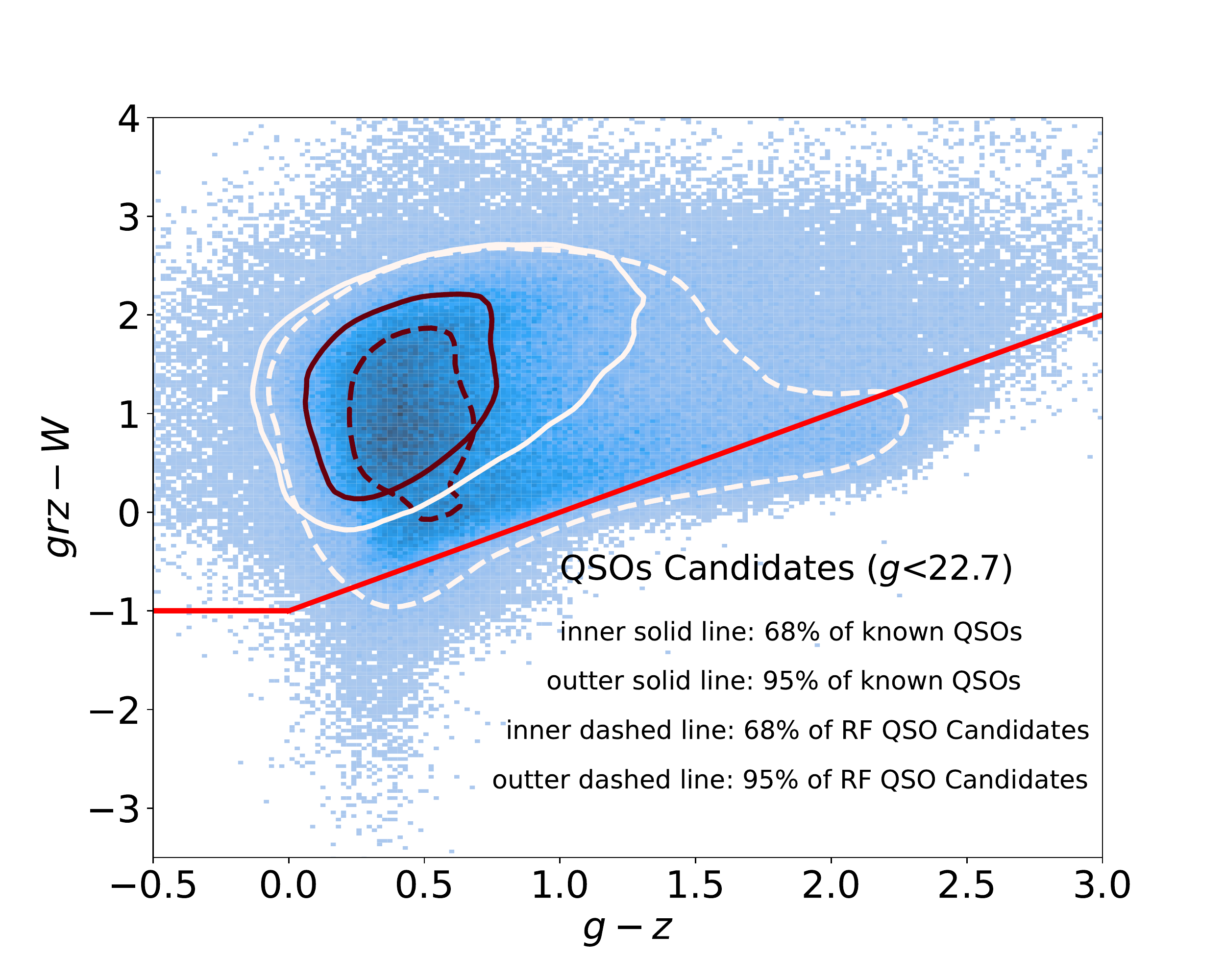}
        \caption{Two-dimensional histogram for comparing RF model and color-cut selection. The definitions of $grz$ and $W$ can be found in Sec.\,\ref{sec:cc} and the red-line represents color-cut condition (Sec.\,\ref{sec:cc}), the points below the line are discarded by color-cut selection.}
        \label{fig:rf__clr}
    \end{figure}

    To cross-check the QSO candidates found by our classifier, we identify QSO candidates using the color-cut selection independently, and the selection criteria are listed in Sec.\,\ref{sec:cc}. Consequently, $8,425,413$ and $19,575,604$ candidates are found by the RF model and color-cut selection separately. There are $6,909,375$ candidates discovered using both approaches, i.e., $\sim 82\%$ RF candidates retrieved by the color-cut selection, as shown in Fig.\,\ref{fig:rf__clr}. The red line represents the hard edge of the color-cut selection, and the color-cut selection discards the RF candidates below this line. The hard-cut leads to $\sim 18\%$ of the RF candidates missed by the color-cut selection because the RF selection gives an irregular shape in the color space. On the other hand, $12,666,119$ ($\sim 65\%$) of color-cut candidates are new compared to RF candidates since the color-cut selection has lower purity than the RF model (see the blue dashed line in Fig.\,\ref{fig:magbin_tst}), bringing in plenty of False Positives.
    
\begin{figure}
        \centering
        \includegraphics[scale=0.35]{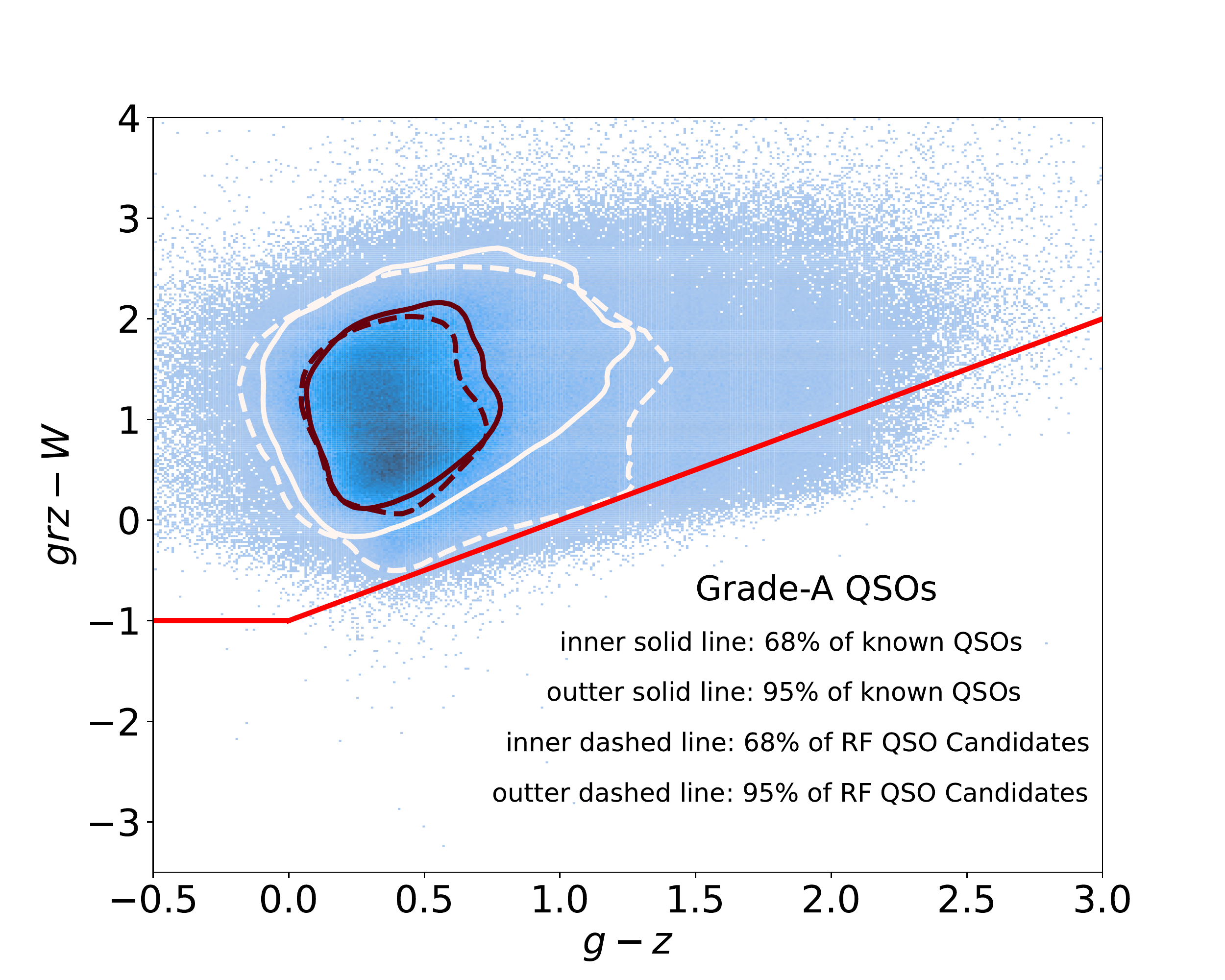}
        \includegraphics[scale=0.35]{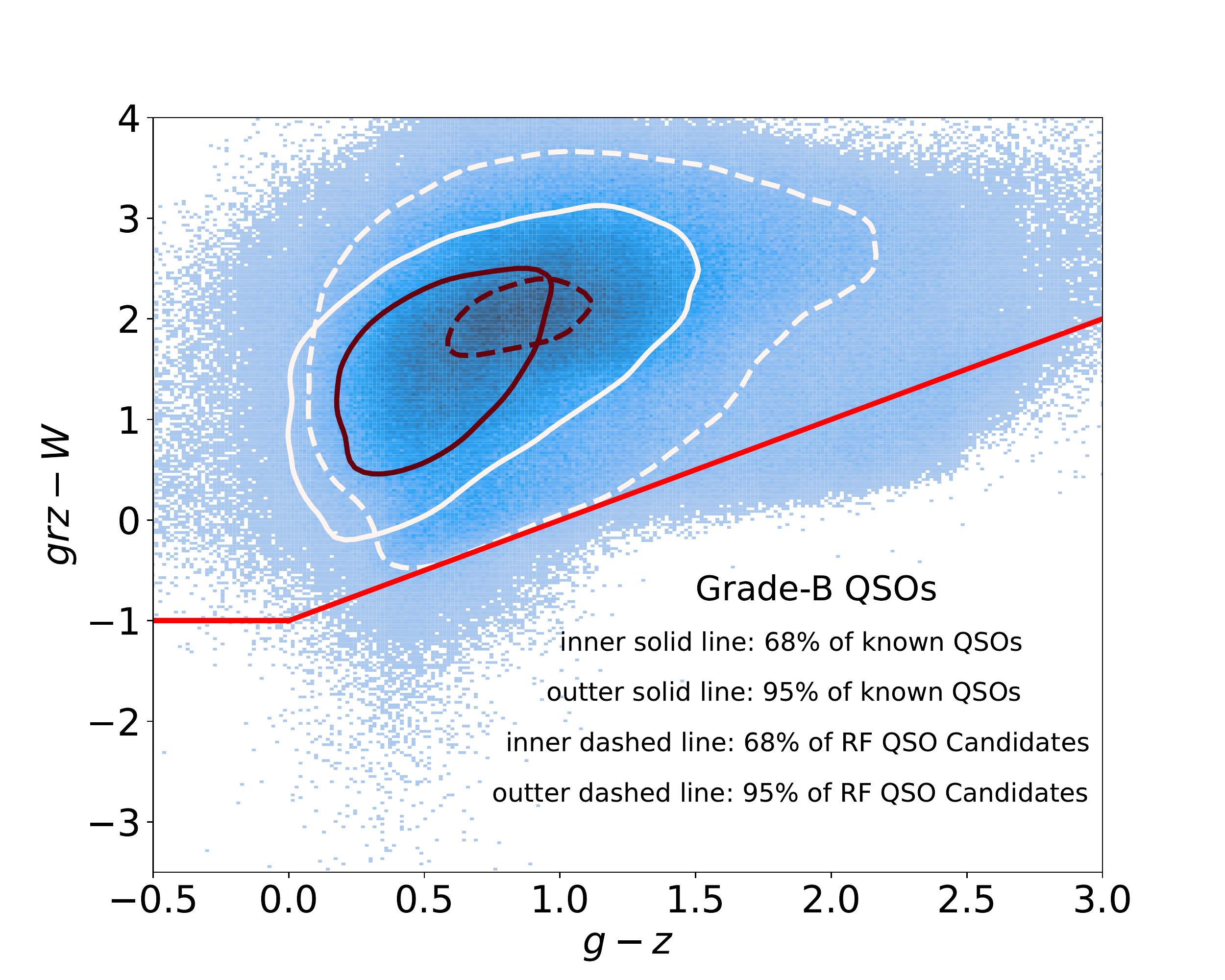}
        \caption{Similar plots with Fig.\,\ref{fig:rf__clr} but for comparing Grade-A candidates (upper) and Grade-B candidates (lower) and the corresponding known quasars.}
        \label{fig:qso_clr_chk}
    \end{figure}

    Furthermore, we test the Grade-A and B candidates in color-cut space as shown in Fig.\,\ref{fig:qso_clr_chk}. Quantitatively, $\sim 91\%$ ($1,771,762$ / $1,953,932$)  Grade-A candidates and $\sim 79\%$ ($5,137,613$ / $6,471,481$) Grade-B candidates and can be re-found by color-cut selection. This result is consist with the ones that could be read from Fig.\,\ref{fig:magbin_tst}: the purity decrease at fainter region.
    
    \begin{table}[]
    \centering
    \begin{tabular}{c|ccc}
    \hline
            & Completeness (TPR) & Purity & FPR     \\ \hline
    A to A  & 1.000        & 0.168  & 0.119  \\ \hline
    B to B  & 1.000        & 0.133  & 0.187  \\ \hline
    AB to A & 0.995        & 0.143  & 0.143  \\ \hline
    AB to B & 0.998        & 0.129  & 0.213  \\ \hline
    \end{tabular}
    \caption{The completeness, purity, FPR of four cases when $0.5$ is adopted as threshold. A represents BASS + MzLS while B represents DECaLS (same with Fig.\,\ref{fig:roc}).}
    \label{tab:whenth0.5}
    \end{table}

\section{Discussion and Summary}\label{sec:dscssn}

    In this work, we have built a catalog of QSO candidates by applying an approach based on Random Forest to the datasets of DESI-LIS and WISE. To train our method, we construct a training set by cross-matching photometry data, including $g$,$r$,$z$ from DESI-LIS and $W1$, $W2$ from WISE, with spectroscopically confirmed QSOs from eBOSS DR16Q for positives and the SIMBAD database for negatives. A testing set mocking the statistical properties of to-be-applied data in magnitude and color is also created and injected to evaluate the completeness, accuracy, and purity of the identification process. At last, $24,440,816$ QSO candidates are identified out of $425,540,269$ point-like objects in DESI-LIS. In addition, we validate our results with those of the color-cut selection approach, and they match well. The catalog can be considered the reference for further observations of DESI and other spectrum surveys to identify new quasars. Relevant data including Grade-A and Grade-B catalogs, training and testing sets are available online\footnote{\url{https://github.com/EigenHermit/he-li2021}}.    Furthermore, the Grade-B candidates in DECaLS are slightly shallower and bluer than those in BASS + MzLS footprints due to the difference in the efficiency of the $g$-band filters in DECaLS and BASS \citep[see Fig.3 in][]{Dey2019}. However, the gap disappears in Grade-A candidates because of their high signal-to-noise ratios. The gaps lead to concerns about the selection of training strategies, i.e., whether the classifier should be trained with the data from the DESI-LIS footprint or only from the DECaLS (or BASS + MzLS) footprint. Our experiments present the completeness is nonsensitive to training strategies when we choose $P_{\rm th} = 0.5$ as the classification threshold (Tab.\,\ref{tab:whenth0.5}). Regardless, additional catalogs of quasar candidates obtained through the RF models trained with the latter training strategy are also published along with the primary catalogs for the others' convenience.
    
    According to the evaluations based on the testing set, the overall purity is $\sim 0.25$ while the completeness is higher than $0.99$. We further define two grades for the candidates by placing a demarcation in $g$-band magnitudes, i.e., the Grade-A catalog contains $1,953,932$ candidates that are brighter than $g=21.18$, while the Grade-B contains $22,486,884$ candidates that are fainter than $g=21.18$. Besides, the accuracy, purity and AUC of the Grade-A catalog are all higher than Grade-B catalog's, specifically, accuracy is $\sim 0.1$ higher, purity is  $\sim 0.15$ higher and AUC is  $\sim 0.05$ higher. However, the completeness of Grade-A is barely the same as Grade-B, which is above 0.85 under all test thresholds (0.5,0.6,0.7,0.8,0.9). The object is considered as a quasar candidate when its RF score is higher than thresholds. We select $0.5$ as the identification threshold for general purposes. With this threshold, $\sim$ 82\% of the quasar candidates found by our method could be rediscovered by the color-cut selection method. Nevertheless, as is expected, a higher threshold leads to lower completeness but higher purity. Thus, one can tweak the threshold to satisfy the requirements of their own scientific goal. 

    We implement the search for QSOs over the whole field of view of DESI-LIS DR9, covering $\sim 14000$ square degrees of the extragalactic sky visible from the northern hemisphere, more extensive than previous work. Besides, by evaluating the classification outcomes of the testing set, we find that the completeness of the QSO candidate catalogs has high completeness when selecting $0.5$ as the identification threshold, which means that the confirmation process with DESI following our targets catalogs can achieve a QSO catalog with both high completeness and purity. However, considering the photometry data adopted in this work ($g, r, z, W1, W2$),  the performance of the identification can be further improved with data in additional bands such as UV and radio. Moreover, blended objects contaminate the photometry catalog due to the PSF size of DESI-LIS; for instance, the objects considered as extended sources are excluded firstly in our work, which might include blended QSOs or blended QSOs and galaxies. Thus, to further increase the completeness of the targets catalog, one needs to conduct a deblending operation over the whole datasets of DESI-LIS before banning negatives, which is part of our further work.
    
    To summarize, this study provides the largest-ever catalog of QSOs candidates with high completeness, which can be used for the targets dataset for confirming QSOs with DESI. Furthermore, grounded on this QSO candidates catalog, we are trying to find the candidates of strongly lensed QSOs with a catalog-based algorithm. So far, $\sim 800$ high-quality candidates of new strongly lensed QSO systems have been found and will be reported in a separate paper. Thorough follow-ups and analyses will be applied to the candidates. Next, we will constrain the properties of the circumgalactic medium \citep{Cai2019,Lau2022}, dark matter distribution in lens galaxies \citep{Oguri2014,Sonnenfeld2021}, Hubble constant \citep{Suyu2017,Liao2019,Wong2020} with confirmed strongly lensed QSO systems.

\begin{acknowledgements}

    We thank the anonymous referee for valuable and constructive comments. We acknowledge the science research grants from the China Manned Space Project with NO.CMS-CSST-2021-A01. We thank China-VO for providing DESI-LIS data and download service for our catalogs. We thank Huanyuan Shan, Rui Li, Dongxu Zhang, Heyang Liu, Jiao Li, Hao Tian, Hui Sun and Jiadong Li for the extensive discussions. We thank astropy, scikit-learn, pandas, seaborn for providing convenient and reliable python packages.

\end{acknowledgements}

\appendix                  

\bibliographystyle{raa}
\bibliography{citation_list}

\label{lastpage}

\end{document}